\documentclass{aa}  
\usepackage{graphicx}
\usepackage{txfonts}
\usepackage{subfig}
\usepackage{multicol}
\usepackage{comment}
\usepackage{float}
\usepackage{booktabs}
\usepackage{colortbl}
\usepackage{siunitx} 
\usepackage[table,xcdraw]{xcolor}
\usepackage{amsfonts, amsmath, amssymb}
\usepackage{mathrsfs}
\usepackage{upgreek}  
\usepackage{newtxtext,newtxmath}
\usepackage{adjustbox}
\usepackage{svg}
\usepackage{soul}

\usepackage[colorlinks=true, citecolor=cyan, linkcolor=blue]{hyperref}
\bibpunct{(}{)}{;}{a}{}{,} % to follow the A&A style
\DeclareUnicodeCharacter{0303}{\~{}} % Tilde character

\begin{document} 
\authorrunning{C. E. Ferreira Lopes et al.}
\titlerunning{Recovering Signals in CoRoT Mission (RSCoRoT)}
\title{Recovering Signals in CoRoT Mission (RSCoRoT):}
\subtitle{I. Short Period Variable Stars}

\author{
  C. E. Ferreira Lopes\inst{1,2},
  A. Papageorgiou\inst{3},
  B. L. Canto Martins\inst{4},
  M. Catelan\inst{5,6,2},
  D. Hazarika\inst{1,2},
  I. C. Leão\inst{4},
  J. R. De Medeiros\inst{4},
  E. Lalounta\inst{3},
  P. E. Christopoulou\inst{3},
  D. O. Fontinele\inst{4}, 
  R. L. Gomes\inst{4}
%  C. E. Ferreira Lopes%\orcid{0000-0002-8525-7977}\inst{1,2},
%  A. Papageorgiou%\orcid{0xxxxx77}\inst{3},
%  B. L. Canto Martins%\orcid{0xxxx7}\inst{4},
%  M. Catelan%\orcid{0000-0001-6003-8877}\inst{5,6,2},
%  D. Hazarika%\orcid{0000-0003-4379-6777}\inst{1,2},
%  I. C. Leão%\orcid{00xxx7}\inst{4},
%  J. R. De Medeiros%\orcid{0xxx77}\inst{4},
%  E. Lalounta%\orcid{00xxx877}\inst{3},
%  P. E. Christopoulou%\orcid{00xxx-8877}\inst{3},
%  D. O. Fontinele%\orcid{0000-xxx}\inst{4}, 
%  R. L. Gomes%\orcid{0000-xxx}\inst{4}
}

\institute{
  Instituto de Astronom\'{i}a y Ciencias Planetarias, Universidad de Atacama, Copayapu 485, Copiap\'{o}, Chile\\
  \email{ferreiralopes1011@gmail.com}
  \and
  Millennium Institute of Astrophysics, Nuncio Monse\~{n}or Sotero Sanz 100, Of. 104, Providencia, Santiago, Chile
  \and
  Department of Physics, University of Patras, 26500, Patra, Greece
  \and
  Departamento de Física Teórica e Experimental, Universidade Federal do Rio Grande do Norte, Campus Universitário, Natal, RN, 59072-970, Brazil
  \and
  Instituto de Astrof\'{i}sica, Pontificia Universidad Cat\'{o}lica de Chile, Av. Vicu\~{n}a Mackenna 4860, 7820436 Macul, Santiago, Chile
  \and
  Centro de Astro-Ingenier\'{i}a, Pontificia Universidad Cat\'{o}lica de Chile, Av. Vicu\~{n}a Mackenna 4860, 7820436 Macul, Santiago, Chile}
\date{Received xxx, 0000; accepted yyy, 0000}

\abstract
{The space mission CoRoT (Convection, Rotation, and planetary Transiting) still holds a wealth of high-quality, yet largely unexplored data that can be analyzed in terms of signal-to-noise ratio and continuous time coverage.}
{This work is the first in a series focused on identifying and classifying variable stars observed by CoRoT whose light curve signals have not yet been studied or reported in the main variable star repositories.}  
{We employed simulations alongside real-world data to assess the effectiveness of the moving average method in handling instrumental jumps and detecting short-period signals (lasting less than one day) within time series spanning approximately 20 days. To classify the newly identified variable stars, we used the light curves of known variable stars as a training set. A supervised selection method was developed, introducing a novel classifier based on features extracted from the folded light curve using the double period.}
{We identified 9,272 variable stars, of which 6,249 are not yet listed in the SIMBAD and VSX repositories. From our preliminary classification, we identified various types of variable stars, including 309 $\beta$ Cephei, 3,105 $\delta$ Scuti, 599 Algol-type eclipsing binaries, 844 $\beta$ Lyrae eclipsing binaries, 497 W Ursae Majoris eclipsing binaries, 1,443 $\gamma$ Doradus, 63 RR Lyrae, and 32 T Tauri stars. The template-based models created serve as a new classifier for variable stars with well-sampled light curves. This catalog introduces CoRoT variable stars into widely used astronomical repositories.} 
{By comparing the properties of the identified variables in the inner and outer regions of the Milky Way, we observed notable differences in several variable star types, likely reflecting metallicity and age gradients. The identification of signals with periods shorter than one day also enables us to propose new approaches for detecting longer-period variability through automated methods, which will be explored in forthcoming papers.}

\maketitle

\section{Introduction}\label{sec:introduction}
The CoRoT mission \citep{Auvergne-2009,Baglin-2007} enabled groundbreaking studies across various astrophysical domains \citep[e.g.,][]{Lanza-2009,Leger-2009,Alencar-2010,FerreiraLopes-2015cycles,Chiappini-2015}. Among its most notable achievements were the discovery of the first Earth-like rocky planet \citep{Leger-2009,Queloz-2009} and the first close-in brown dwarf with a measured radius \citep{Bouchy-2011A,Csizmadia-2015}. The high-quality, homogeneous photometric data from CoRoT have enabled significant advancements in the study of variable stars, including RR~Lyrae stars \citep{Chadid-2010}, eclipsing binaries \citep{Dolez-2009, Maciel-2011}, and rotational and pulsating variable stars \citep{DeMedeiros-2013}. Key findings include the identification of 4,106 CoRoT sources with semi-sinusoidal signatures \citep{DeMedeiros-2013}, a catalog of 1,978 stars with periodicities \citep{Affer-2012}, and the discovery of 418 $\gamma$~Doradus, 274 hybrid $\gamma$~Doradus/$\delta$~Scuti candidates \citep{Hareter-2012}, as well as 1,428 pulsating M giants \citep{FerreiraLopes-2015mgiant}. Additionally, 74 $\delta$~Scuti and hybrid stars were analyzed to advance the understanding of A--F-type stars \citep{Moya-2017}. CoRoT data have also contributed significantly to the study of stellar magnetic activity cycles. For example, \citet{Garcia-2010} identified a solar-like activity cycle in the main-sequence star HD~49933 using asteroseismic analysis of CoRoT light curves. Similarly, \citet{FerreiraLopes-2015cycles} reported potential new activity cycles in a sample of 16 FGK main-sequence stars observed by CoRoT.

\begin{figure*}
\centering
\includegraphics[width=0.48\textwidth]{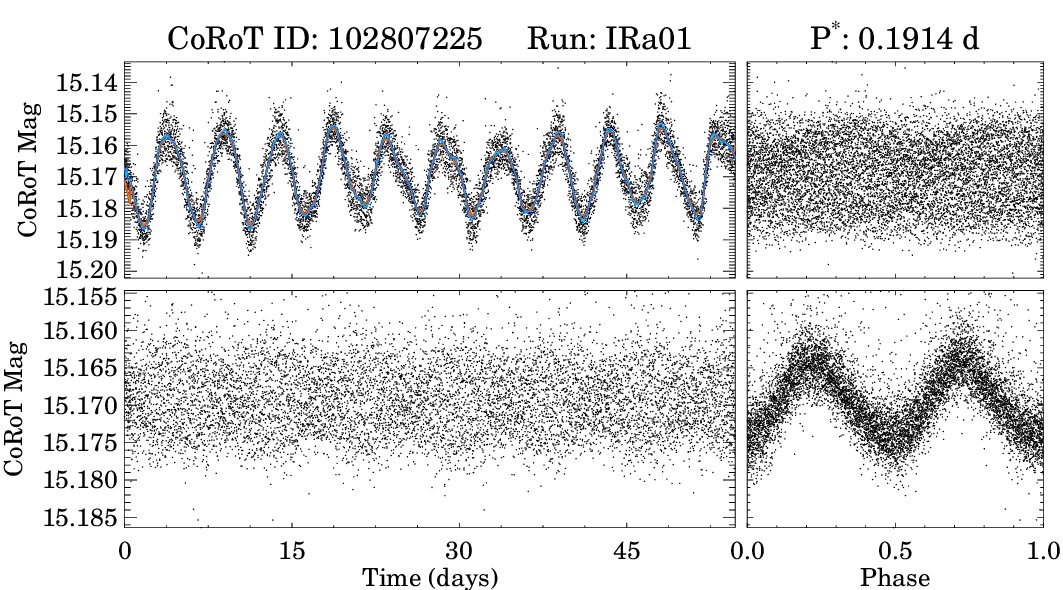}
\includegraphics[width=0.48\textwidth]{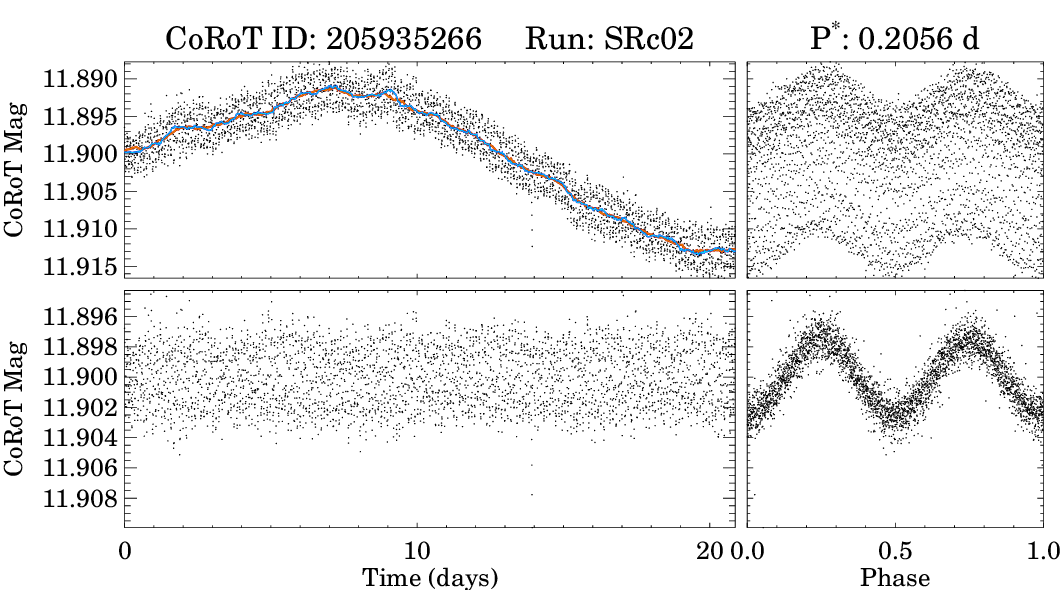} \hspace{1cm}
\includegraphics[width=0.48\textwidth]{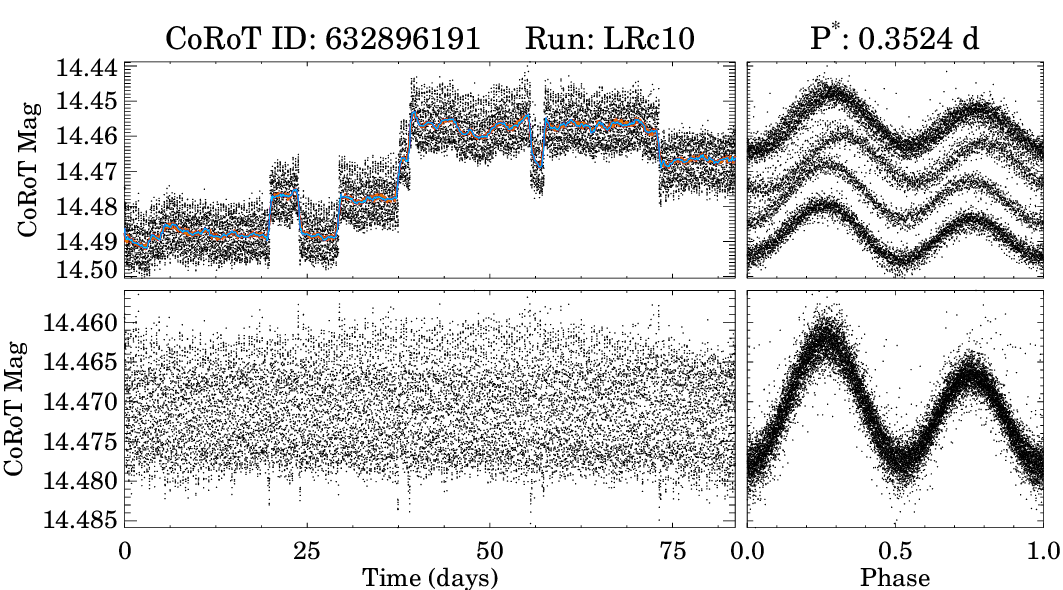}
\includegraphics[width=0.48\textwidth]{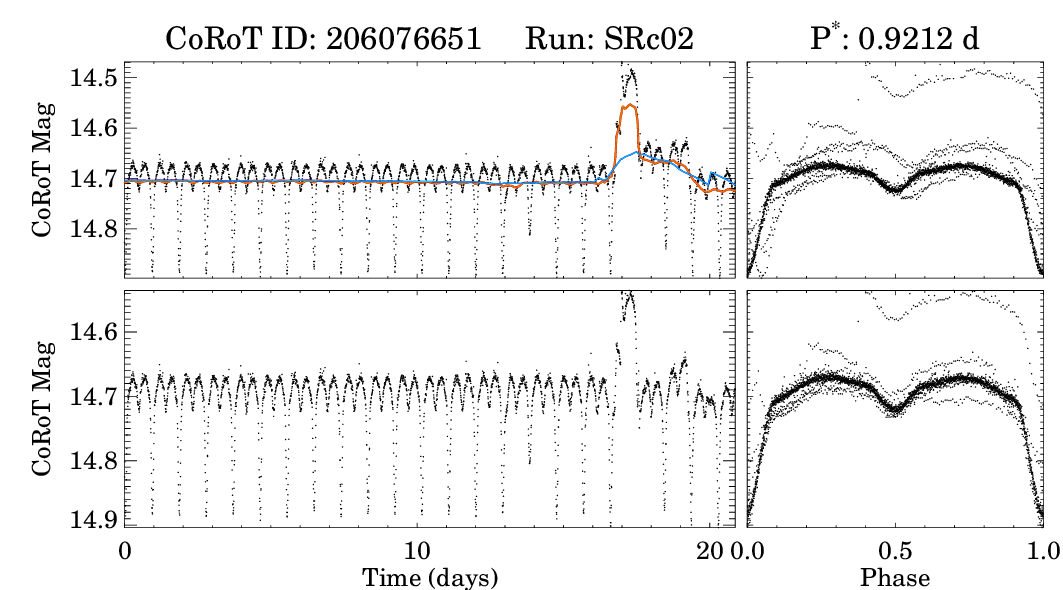}
\caption{CoRoT light curves before (upper panels) and after (bottom panels) applying the Moving Average Method (MAM).Phase-folded light curves are shown in the top-right corner of each panel. The orange line represents the MAM with a time segment size (TSS) of 1 day, while the blue line corresponds to a TSS equal to the variability period. Each panel is labeled with the CoRoT ID, field, and period of variability at the top.}
\label{fig_CorotLCs}
\end{figure*}

However, a comprehensive catalog of variable stars with well-defined CoRoT light curves is still lacking. Indeed, catalogs containing a large number of false classifications or poorly selected sources are typically excluded from major variable star repositories such as \textsc{SIMBAD} or the VSX database, which prioritize reliable and well-vetted data. For instance, \citet{FerreiraLopes-2020-VIVA} conducted a comprehensive selection of variable star candidates using multiple variability indices, producing a catalog of 45 million candidates. However, many of these candidates likely suffer from poor-quality signals, particularly for very faint or very bright stars that often lack trustworthy photometric variability. Similarly, \citet{Debosscher-2007,Debosscher-2009} classified CoRoT data using Fourier models of the light curves, but shifts in baseline levels often led to unreliable parameters and false classifications. Although the authors provided probability flags to improve target selection, a complete list of high-quality variable stars was not made publicly available. Likewise, \citet[][]{Gavras-2023} compiled a comprehensive catalog of variable stars cross-matched with Gaia data; however, only about 22\% of the variable stars identified in this work are included in that compilation.

The CoRoT database has also served as a testbed for developing new methods to extract and classify information from stellar light curves (LCs) \citep{FerreiraLopes-2018papIII}. For instance, \citet{Damiani-2016} provided the first spectral and luminosity classification of CoRoT targets using broadband multi-color photometry. Spectroscopic data have further enhanced the characterization of CoRoT targets \citep[e.g.,][]{Guenther-2012,Sarro-2013,Anders-2017,Anders-2020}. However, the reliability of such classifications can be compromised if the light curves are affected by instrumental biases or discontinuities. To address this, various algorithms have been developed to correct CoRoT LCs \citep[e.g.,][]{Mislis-2010,DeMedeiros-2013,Leao-2015}, significantly improving data quality by minimizing jumps, trends, and outliers. These improvements have enhanced the statistical robustness of the data, enabling more accurate confidence intervals and reducing over- or underestimation errors, yet a complete catalog of CoRoT sources with reliable signals has not been produced.

These improvements have opened new avenues for exploring CoRoT data in different science cases, including the global characterization of faint Be stars in the CoRoT exofield \citep[][]{Zorec-2023}, eclipsing binaries exhibiting the light-travel-time effect \citep[][]{Hajdu-2022}, the detection and characterization of planets around intermediate-mass stars \citep[][]{Sebastian-2022}, and period fluctuations in CoRoT RR~Lyrae stars \citep[][]{Benko-2016}. However, the selection and characterization of variable star candidates observed by CoRoT remain incomplete. Challenges arise when authors rely on such catalogs to select targets but fail to publish them, hindering efforts to compile and validate variable star discoveries. This lack of accessibility presents significant obstacles for cross-matching and validating classifications.

This work is the first in a series within the \textit{New Insight Into Time Series Analysis} (NITSA) project, which aims to enhance time-series analysis techniques—particularly in the presence of level shifts—and to construct a comprehensive catalog of variable stars with reliable signals from CoRoT data \citep{FerreiraLopes-2016papI,FerreiraLopes-2017papII,FerreiraLopes-2018papIII,FerreiraLopes-2020-VIVA,FerreiraLopes-2021papIV}. To this end, we use the latest version of the CoRoT database \citep[][]{Chaintreuil-2016,Ollivier-2016}, which incorporates significant improvements over earlier releases, notably in mitigating instrumental jumps caused by temperature fluctuations or proton impacts on the CCDs. Despite these advances, a full selection and characterization of the CoRoT variable star population has not yet been completed.

The paper is organized as follows: Section~\ref{sec_dataset} introduces the Moving Average Method (MAM) as a tool to address data quality issues. Section~\ref{sec_cat} describes the selection process for variable stars. Sections~\ref{sec_cross} and \ref{sec_debosher} detail the cross-matching with external catalogs and the classification of new targets. Section~\ref{sec_results} presents the main findings, and Section~\ref{sec_conclusions} provides a summary of our conclusions.

\section{CoRoT dataset}\label{sec_dataset}	

The CoRoT telescope observed 163,665 point sources distributed across 26 stellar fields in the faint star channel. Among these, 12,896 sources were observed in two separate campaigns, while 892 sources were observed in three. The first CoRoT field (IRa01) was monitored for 54.3 days, while the remaining fields were divided into short (S) and long (L) runs, with durations ranging from 5 to 52.3 days and 76.7 to 148.3 days, respectively. Observations were conducted with a cadence of either 512 seconds or 32 seconds, the latter being used for select oversampled targets \citep[e.g.,][]{Auvergne-2009}. The survey also provided luminosity classes and spectral types for the observed stars, determined based on their positions in the color--magnitude diagram \citep[][]{Damiani-2016,Deleuil-2009,Deleuil-2018}. Notably, the SRc03 CoRoT run had the shortest time coverage, approximately 3.5 days, but around 85\% of its stars were re-observed in subsequent runs.

A shift-level variation in a light curve refers to sudden changes in its mean value, typically caused by abrupt drops or increases in pixel sensitivity. These variations are often associated with instrumental issues or the impact of high-energy particles on the CCD \citep[e.g.,][]{Auvergne-2009}. In CoRoT data, most perturbations arise from high-energy particle interactions with the CCD (see, for example, the raw light curves in Fig.~\ref{fig_CorotLCs}). To address this issue, we applied the Moving Average Method (MAM), which computes averages over subsets of data within a defined time segment size (TSS). For evenly spaced data, the TSS parameter corresponds to the number of data points used in the calculation of the moving average. MAM acts as a prewhitening filter, attenuating signals with periods longer than the TSS (see Fig.~\ref{fig_CorotLCs}) and effectively mitigating multiple jumps, outliers, and long-term trends (e.g., CoRoT 632896191 and 206076651 in Fig.~\ref{fig_CorotLCs}). However, it does not specifically correct for abrupt shifts, which can introduce biases in segments of data with lengths comparable to the TSS.

Examples of CoRoT signals before (upper panels) and after (lower panels) applying MAM corrections are shown in Fig.~\ref{fig_CorotLCs}. The effectiveness of the correction is particularly evident in long-period signals, such as for source 206076651, where the period is significantly longer than those of other targets. Additionally, the signal removed by MAM may correspond to long-term trends, contamination from background stars \citep[e.g.,][]{FerreiraLopes-2015mgiant}, or variability with periods longer than the TSS (see sources 102807225 and 205935266 in Fig.~\ref{fig_CorotLCs}). This study includes the full CoRoT dataset, except the SRc03 subset, which was excluded due to its limited time coverage.

The MAM is a well-established technique that is straightforward to understand and implement. It effectively reduces biases in CoRoT data, aiding in the recovery of short-period signals (see Sect.~\ref{sec_TestSimulations}). To assess the potential biases introduced by MAM in CoRoT light curves, we conducted a series of simulations using representative signal types found in CoRoT data. These simulations enabled us to quantify the extent to which MAM modifies signal parameters and to evaluate its impact on the extracted light curve characteristics.

\begin{figure}
  \centering
  \includegraphics[width=0.48\textwidth, height=0.7\textwidth]{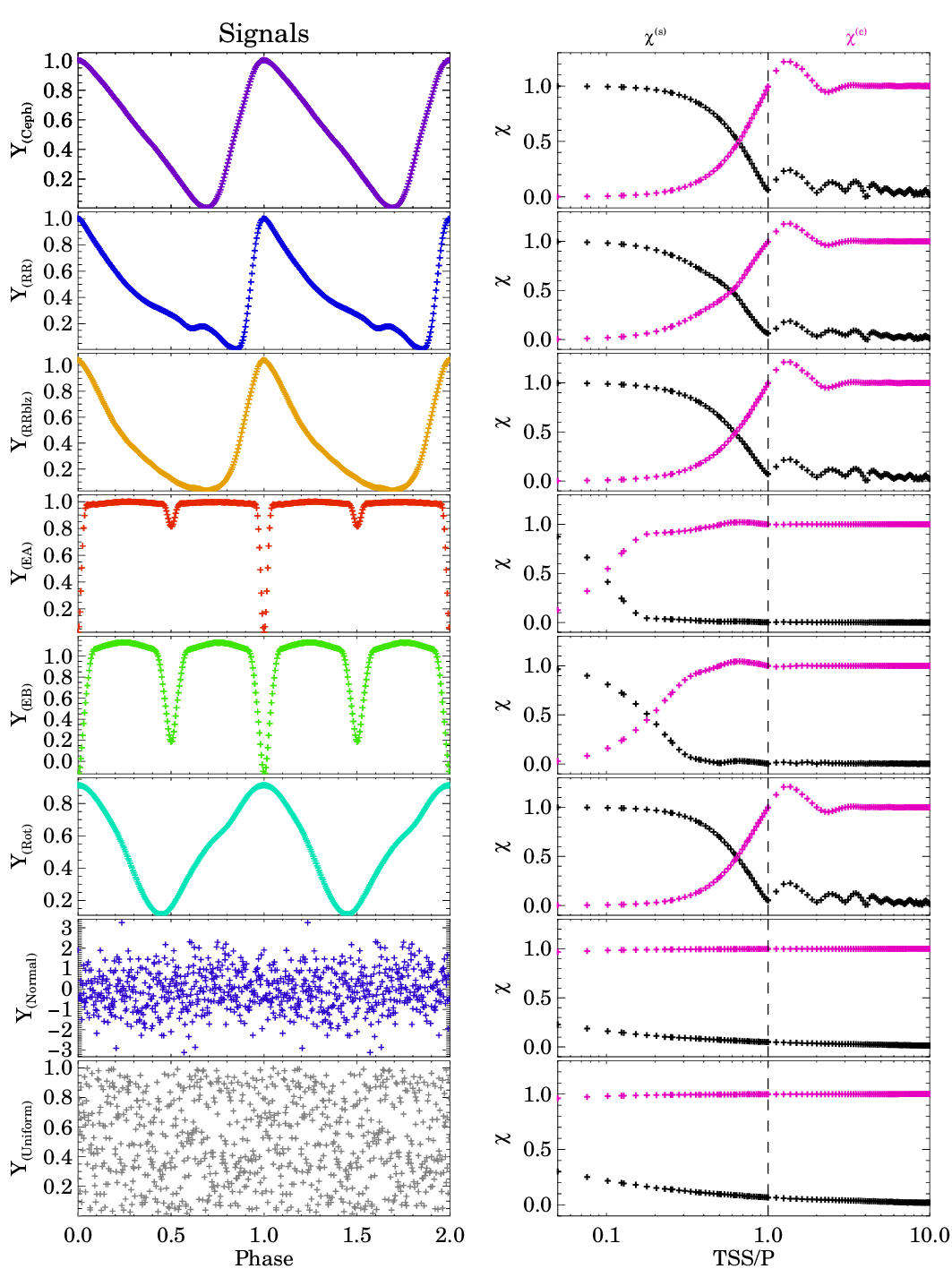}
   \caption{Folded CoRoT signals (left panels) and $ \chi^{(c)}$ and  $\chi^{(s)}$ parameters (see Eqs. \ref{eq_par01} and \ref{eq_par02}) as a function of the ratio between the $TSS$ length and the period. The dashed black line sets  the position where $\rm TSS = P$, i.e. $\rm TSS/P = 1$.}
  \label{fig_SimCorot}
\end{figure}

\subsection{Design of the simulations}\label{sec_TestSimulations}	

To assess the impact of the smoothing process on time-series data, we generated synthetic light curves without noise or artificial jumps. A comprehensive simulation typically accounts for all possible variables, including jump amplitude, number of occurrences, signal phase, and the data behavior before, during, and after each jump, as well as overlay complexity. Therefore, we opted for a more straightforward simulation approach. These simulations were designed to isolate the effect of the moving average and evaluate how the smoothing window (with a length of TSS) interacts with the intrinsic period of the signal (P). In this framework, the period becomes a dependent variable concerning TSS, and the key quantity is the ratio TSS/P. Therefore, the specific choice of period is not crucial, as long as the comparison is made between signals with equivalent TSS/P ratios. Since the moving average filter only modifies the data locally over segments of length TSS, it primarily affects features with spatial scales comparable to or larger than TSS. The results obtained using the ratio between \(\rm TSS\) length and period can be generalized to any period value. To quantify how \(\rm TSS\) modifies the signal, we used the following parameters:

\begin{equation}
  \chi^{(c)} = \frac{  \sum_{i=1}^{i=N} \left|  y_i^{(o)} - y_i^{(f)} \right| }{ \sum_{i=1}^{i=N} \left| y_i^{(o)} \right| }  =  \frac{  \sum_{i=1}^{i=N} \left|  y_i^{(c)} \right| }{ \sum_{i=1}^{i=N} \left| y_i^{(o)} \right| }  
  \label{eq_par01}
\end{equation} 

\noindent and

\begin{equation}
  \chi^{(s)} = \frac{ \sigma^{(s)} }{ \sigma^{(o)} }, 
  \label{eq_par02}
\end{equation} 

\noindent where \( y_i^{(f)} \) is the fit obtained using the MAM, \( y_i^{(o)} \) is the original data (measured in fluxes or magnitudes), \( \sigma^{(o)} \) is the standard deviation of \( y_i^{(o)} \), \( y_i^{(c)} \) is the corrected data (i.e., \( y_i^{(o)} - s_i \)), and \( s_i \) is the correction factor calculated from the MAM (see red lines in Fig. \ref{fig_CorotLCs}). Equation~\ref{eq_par01} defines a normalized cumulative absolute deviation between the original and filtered signals. This metric is scale-invariant and captures the impact of the filtering process as a fraction of the original signal’s magnitude, allowing for direct comparison across light curves with different amplitudes. In contrast, Equation~\ref{eq_par02} represents the ratio of standard deviations before and after applying the smoothing. This quantity provides a direct measure of the variability suppressed by the moving average, making it a valuable diagnostic for evaluating the influence of filtering on the time-domain structure of the signal.

The rate at which the distortion metrics $\chi^{(c)}$ and $\chi^{(s)}$ approach their asymptotic values is governed principally by the \emph{duty cycle}\footnote{The fraction of time that a signal or phenomenon is active or observable during a given period.} of the variable‐star signal. For example,  detached and semi‐detached systems the flux is nearly constant for most of the orbital cycle (see Fig. \ref{fig_SimCorot} \(Y_{\rm (EA)}\) and \(Y_{\rm (EB)}\)), because eclipse events last only a small fraction $d = \Delta t_{\mathrm{ecl}}/P\lesssim0.05$–$0.15$ \citep[e.g.][]{Prsa-2011,Maciel-2011,Carmo2020}. When the smoothing window satisfies $\mathrm{TSS}\gtrsim dP$, almost every position of the window is dominated by out‐of‐eclipse samples.  Consequently the moving average hardly alters the light curve and the cumulative‐correction ratio $\chi^{(c)}$ drops rapidly towards zero, while the standard‐deviation ratio $\chi^{(s)}$ tends to unity. On the other hand, pulsating stars, spotted rotators, and ellipsoidal variables \citep[e.g.][]{FerreiraLopes-2015cycles,FerreiraLopes-2015mgiant,FerreiraLopes-2015wfcam, FerreiraLopes-2021papIV,Baeza-Villagra-2025}
exhibit brightness changes throughout the entire period (see Fig. \ref{fig_SimCorot} \(Y_{\rm (Ceph)}\), \(Y_{\rm (RR)}\), \(Y_{\rm (RRblz)}\) and \(Y_{\rm (Rot)}\)). In this case, every placement of the window mixes epochs of different flux, so the filter removes power over a broader Fourier range.  The convergence of $\chi^{(c)}$ is therefore slower, and $\chi^{(s)}$ decreases monotonically with $\mathrm{TSS}/P$ until it reaches the white‐noise floor.

\subsection{Simulation results}

We conducted \(10^6\) simulations to investigate the correlation between the \(\rm TSS\) factor and modifications in the time series signal. The simulations considered typical CoRoT signals representing rotating variable stars (\(Y_{\rm (Rot)}\)), Beta Lyrae eclipsing binaries (\(Y_{\rm (EB)}\)), Algol eclipsing binaries (\(Y_{\rm (EA)}\)), pulsating stars (\(Y_{\rm (Ceph)}\), \(Y_{\rm (RR)}\), \(Y_{\rm (RRblz)}\)), and white noise (\(Y_{\rm (Uniform)}\) and \(Y_{\rm (Normal)}\)), as shown in the left panels of Figure \ref{fig_SimCorot} \citep[see][for more details]{FerreiraLopes-2018papIII}. The synthetic signals were generated using sinusoidal functions with representative profiles for standard classes of variable stars, with periods spanning a range adequate to cover TSS/P ratios from below $0.05$ to $10$. The cadence and total time span of the simulations were consistent with typical CoRoT cadence space-based photometry but without introducing observational gaps or phase discontinuities. No CoRoT-like noise or instrumental effects were added, allowing us to isolate the mathematical impact of the smoothing process.  Key findings from our simulations are as follows (see Figure \ref{fig_SimCorot}):

\begin{itemize}
  \item The bias introduced by the MAM, computed over segments of length \(\rm TSS\), on a time series with period shorter than \(\rm TSS\) (i.e., \(\rm TSS/P > 1\)) decreases as the \(\rm TSS/P\) ratio increases for all non-noise signals analyzed. This trend is observed through the behavior of the parameters \( \chi^{(c)} \) and \( \chi^{(s)} \), which tend toward 1 and 0, respectively. These results demonstrate that increasing the segment length effectively reduces the bias introduced by the moving average method.

  \item When \(\rm TSS/P > 1\), only small variations in the parameters \( \chi^{(c)} \) and \( \chi^{(s)} \) are observed, particularly for signals such as \( Y_{(\rm Ceph)} \), \( Y_{(\rm RR)} \), \( Y_{(\rm RRblz)} \), and \( Y_{(\rm Rot)} \).

  \item For noisy data, represented by normal and uniform distributions, the values of \( \chi^{(c)} \) and \( \chi^{(s)} \) remain approximately constant. This indicates that the MAM does not significantly affect the statistical properties of noise-dominated light curves.

  \item The rapid increase of \( \chi^{(c)} \rightarrow 1 \) and \( \chi^{(s)} \rightarrow 0 \) occurs more quickly for signals such as \( Y_{(\rm EA)} \) and \( Y_{(\rm EB)} \). This behavior is explained by the quasi-constant flux outside the eclipses, where a larger number of points contribute to the moving average, particularly at higher \(\rm TSS/P\) values.
\end{itemize}
 
In summary, the MAM introduces only a minor modification to the signal when \(\rm TSS/P > 1\), while the characteristics of noisy data remain largely unaffected across the entire range of \(\rm TSS/P\). With \(\rm TSS = 1\), we can effectively analyze CoRoT signals with periods shorter than one day. Since both short (SR) and long (LR) CoRoT runs typically span more than 20 days, this further supports the use of \(\rm TSS = 1\) to access and analyze the full CoRoT dataset.

Each data point in the MAM calculation incorporates measurements within the \(\rm TSS\) length. For a light curve with a duration of 20 days, approximately 10\% of the data points computed by the MAM may exhibit bias in the case of a single jump. However, when additional observations before and after a jump are included in the calculation, the resulting averages become less biased and more representative of the true underlying signal (see lower right panel in Fig.~\ref{fig_CorotLCs}). Consequently, the percentage of biased data points is expected to be lower in realistic scenarios involving multiple data segments.

It is important to note that signals with asymmetries exhibit smaller variations compared to the main signals, as shown in Fig.~\ref{fig_SimCorot}, and thus require more detailed analysis. Future papers in this series will investigate alternative methodologies for characterizing variable stars with longer periods.

\begin{figure}
  \centering
  \includegraphics[width=0.5\textwidth]{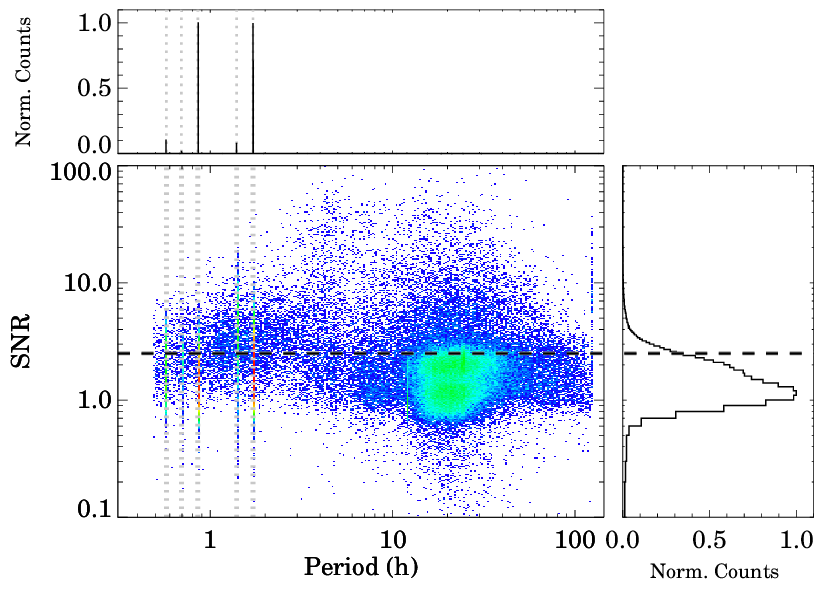}
  \caption{Distribution of the signal-to-noise ratio (SNR) as a function of the period. The horizontal black line indicates the adopted threshold (\(\rm SNR > 2.5\)) used to select variable star candidates. Histograms in the top and right panels show the marginal distributions of period and SNR, respectively. Additionally, grey dotted lines mark concentrations of periods at specific values, indicating possible instrumental or sampling artifacts, such as, thermal cycling, scattered light, and satellite observing cadence.}
  \label{fig_SNRPeriod}
\end{figure}

In this study, a one-day moving average was applied to the entire CoRoT dataset, and the generalized Lomb-Scargle periodogram \citep[LSG;][]{Lomb-1976,Scargle-1982,Zechmeister-2009} was used to search for variability periods. The signal-to-noise ratio \citep[e.g.,][]{DeMedeiros-2013,FerreiraLopes-2015cycles} was then computed to identify potential variable star candidates, as described in subsequent sections and illustrated in Fig.~\ref{fig_SNRPeriod}.

\section{Sample selection and initial classification}\label{sec_cat}	

Our initial sample was selected based on two parameters: the signal-to-noise ratio (SNR) and a period bias flag. We used a bin of \( \pm 10^{-7} \) days to estimate the number of stars sharing similar periods. If more than 10 stars fell within this interval, the corresponding period was flagged as biased. This flag is listed as \( \rm Flag(Period) \) in the catalog (see Table~\ref{table_data}). Based on these criteria, we obtained a sample of 10,196 stars with SNR \(> 2.5\) and non-biased period, as shown in Fig.~\ref{fig_SNRPeriod}.

To simplify the visual inspection process, we propose a supervised selection method called the Light Curve Shape Selection Method (LC-SSM). This method groups similar signals based on a template model, where signals within a given \( \chi^2 \) threshold are considered to have similar shapes. The LC-SSM requires two free parameters: the minimum \( \chi^2 \) value and the number of sources per template model (\( N_{S} \)). However, it can be reduced to a single parameter by defining the minimum \( \chi^2 \) value as a function of the number of sources per template model. The LC-SSM is applied using the following procedure:

To simplify the visual inspection process, we propose a supervised selection method called the LC Shape Selection Method (LC-SSM). This method groups similar signals based on a template model, which serves as a reference light curve. Signals with a reduced chi-squared (\( \chi^2 \)) below a defined threshold relative to the template model are considered to have similar shapes.  The LC-SSM requires two free parameters: the minimum acceptable \( \chi^2 \) value and the number of sources (\( N_S \)) assigned to each template model. However, this can be reduced to a single parameter by determining the \( \chi^2 \) threshold as a function of \( N_S \). The LC-SSM is implemented using the following procedure, as illustrated in the Fig. \ref{fig:workflow}.:

\begin{figure}[ht]
    \centering
    \includegraphics[width=0.75\linewidth]{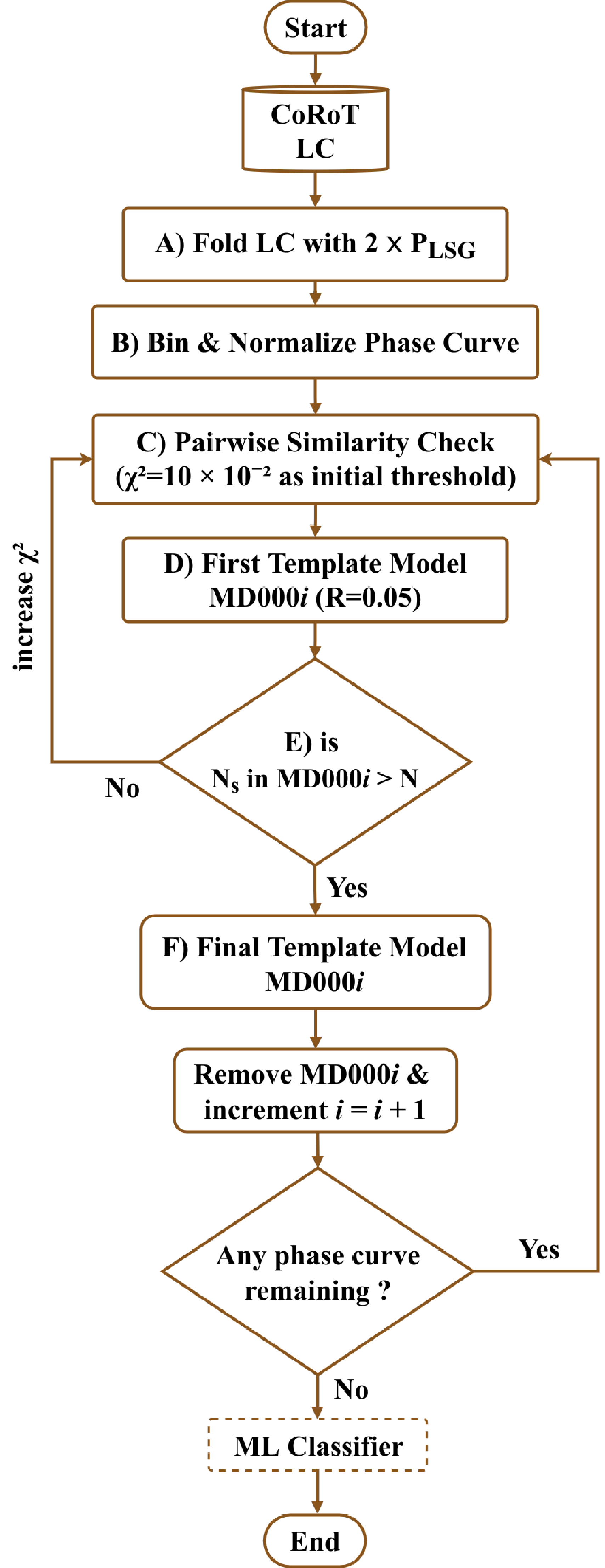}
    \caption{Workflow diagram of the LC-SSM proposed in this study.}
    \label{fig:workflow}
\end{figure}

\begin{itemize}
    \item A – The light curve (LC) is folded using twice the period obtained from the LSG method (see Section~\ref{sec_cross} for more details).
    
    \item B – The phase template model, labeled MD0001. Although harmonic fits could be used, binning the phase diagram is faster and sufficiently accurate for well-sampled data. Step C is then repeated to assign all remaining sources that match MD0001.
    
    \item E – If the number of sources assigned to MD0001 exceeds the user-defined threshold (\( N_{S} \)), it is retained as a template model. Otherwise, the \( \chi^2 \) threshold is incremented, and steps C–D–E are repeated.
    
    \item F – This procedure is repeated until a maximum \( \chi^2 \) threshold is reached or the number of unassigned sources falls below a minimum threshold.
\end{itemize}

\begin{figure*}
  \centering
  \includegraphics[width=0.33\textwidth]{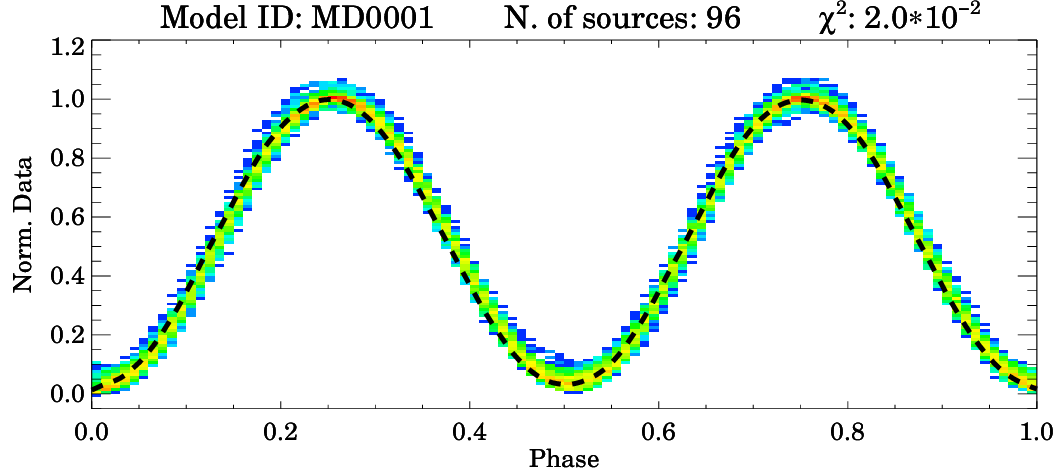}
  \includegraphics[width=0.33\textwidth]{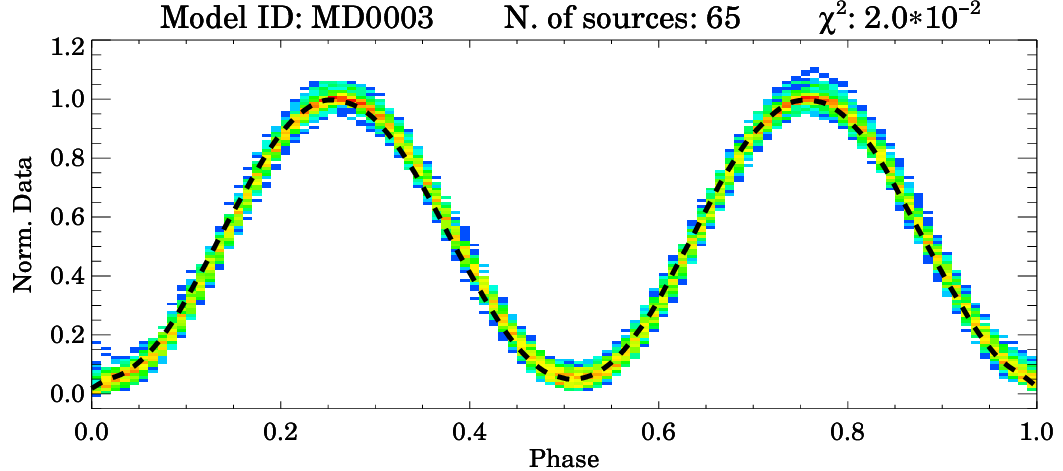}
  \includegraphics[width=0.33\textwidth]{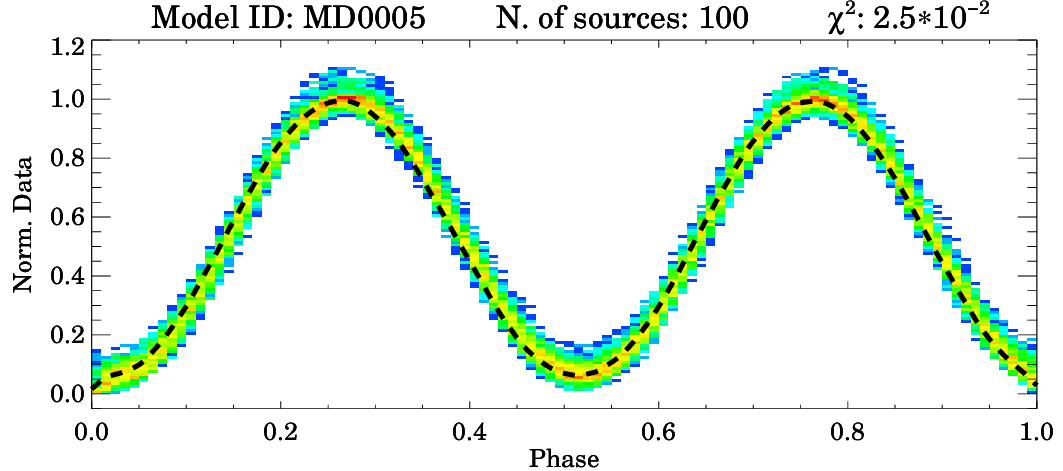}
  
  \includegraphics[width=0.33\textwidth]{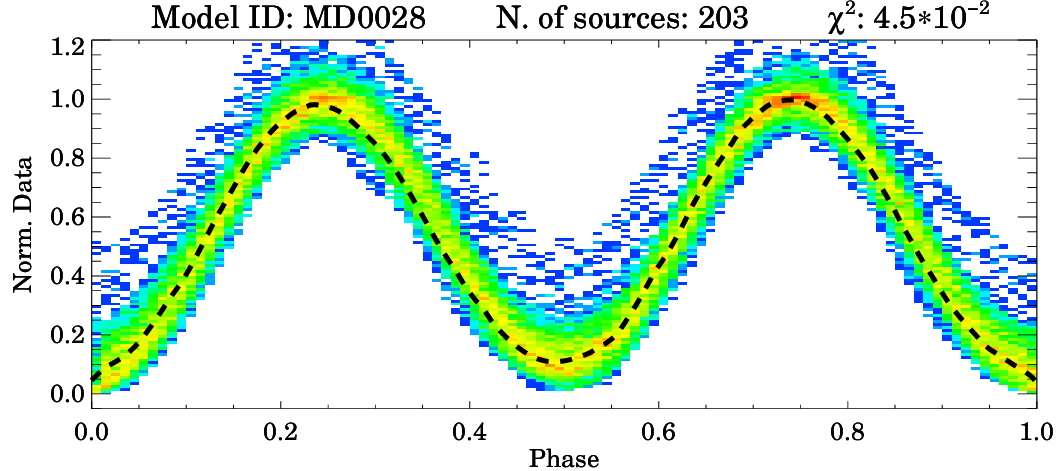}
  \includegraphics[width=0.33\textwidth]{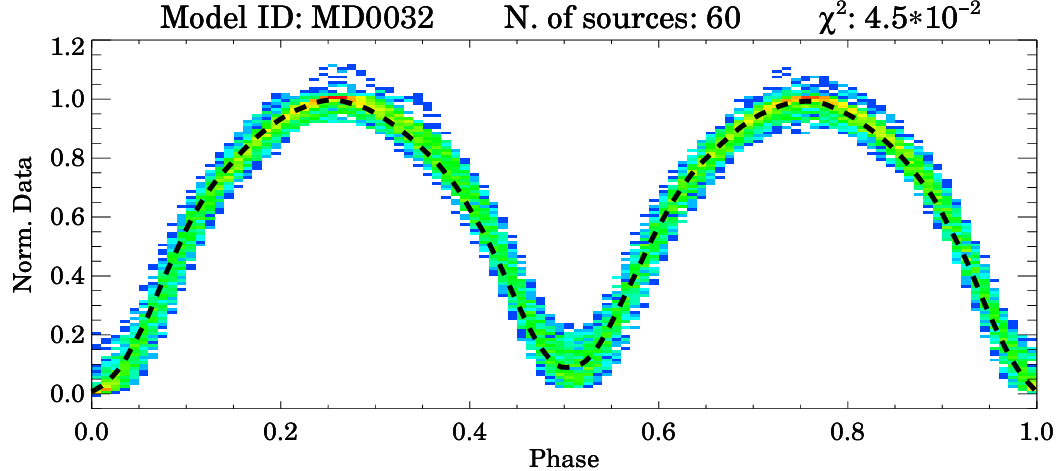}
  \includegraphics[width=0.33\textwidth]{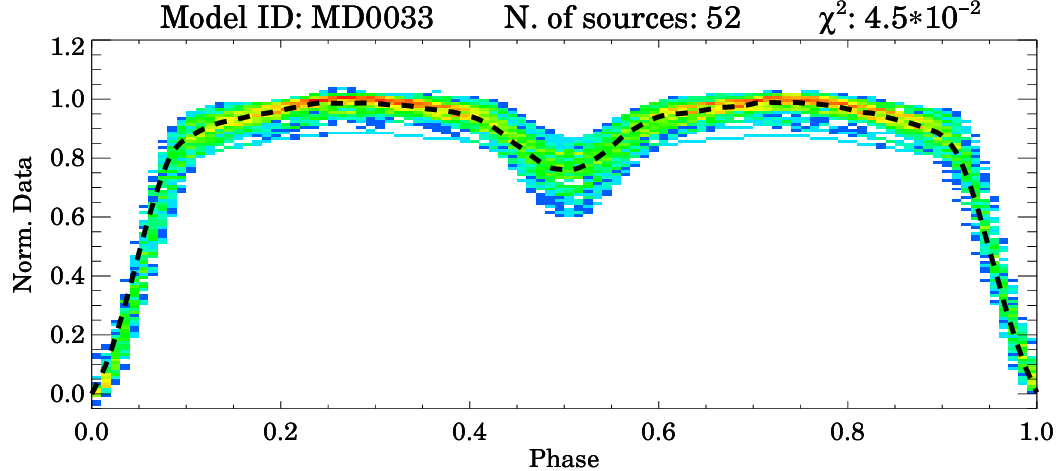}

  \includegraphics[width=0.33\textwidth]{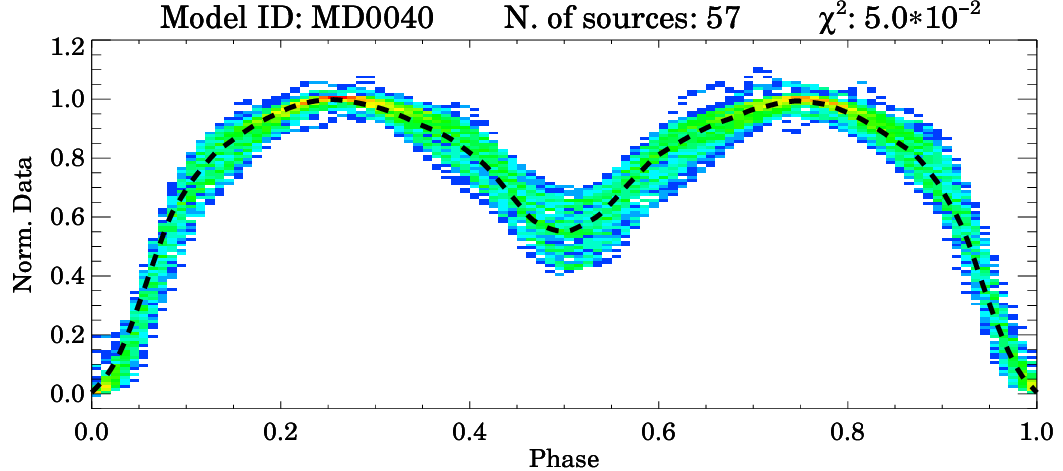}
  \includegraphics[width=0.33\textwidth]{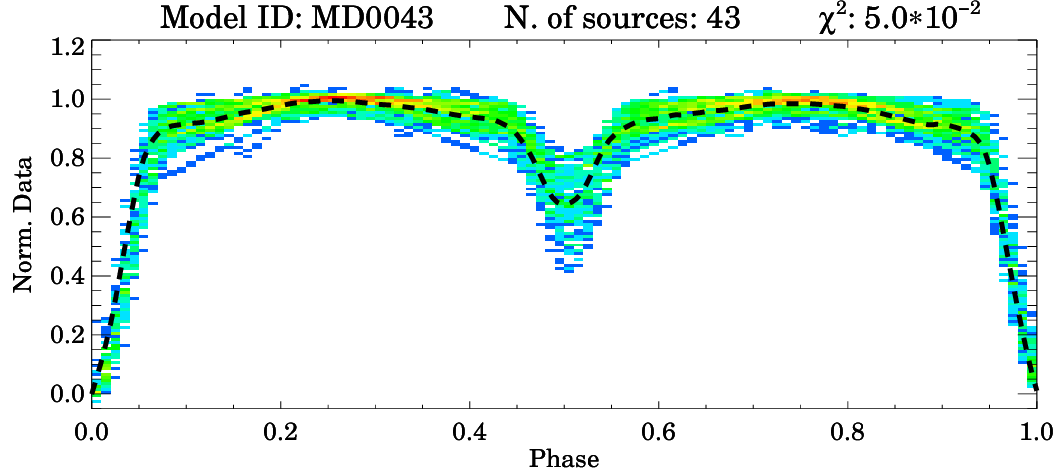}
  \includegraphics[width=0.33\textwidth]{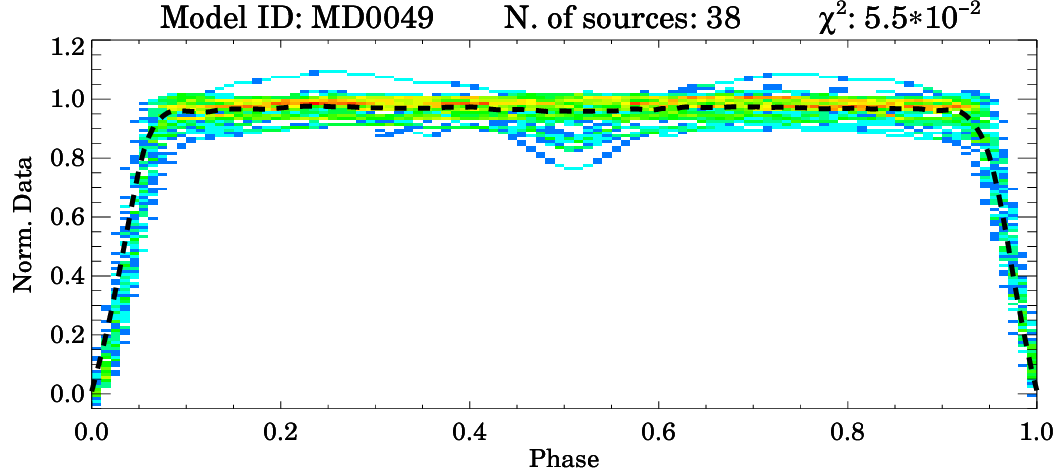}
  
  \includegraphics[width=0.33\textwidth]{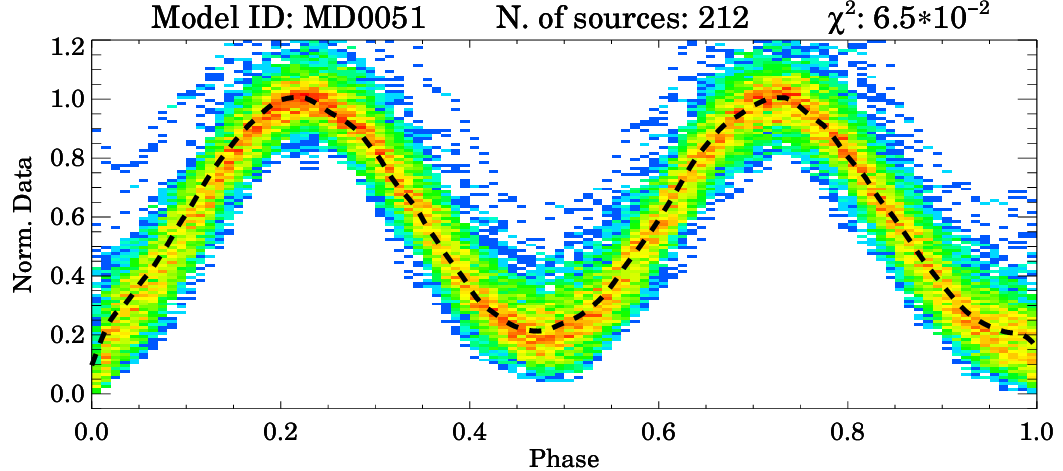}
  \includegraphics[width=0.33\textwidth]{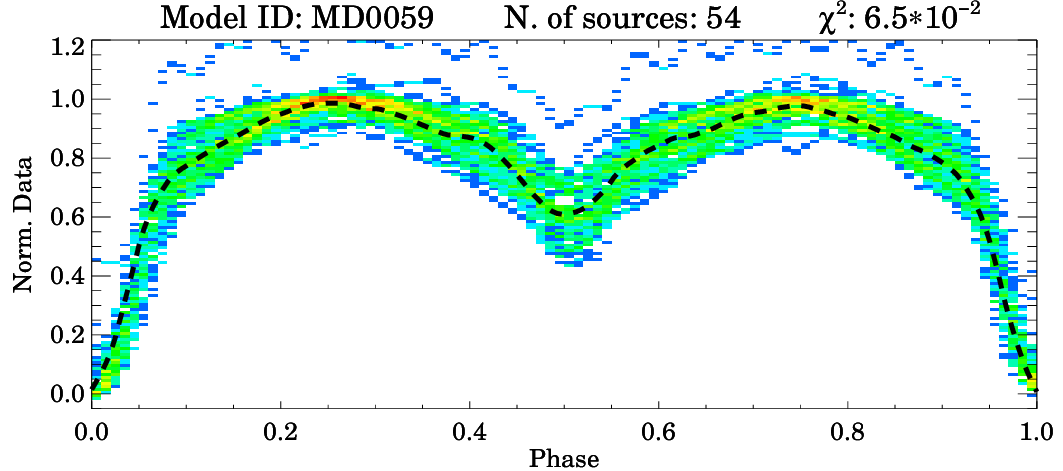}
  \includegraphics[width=0.33\textwidth]{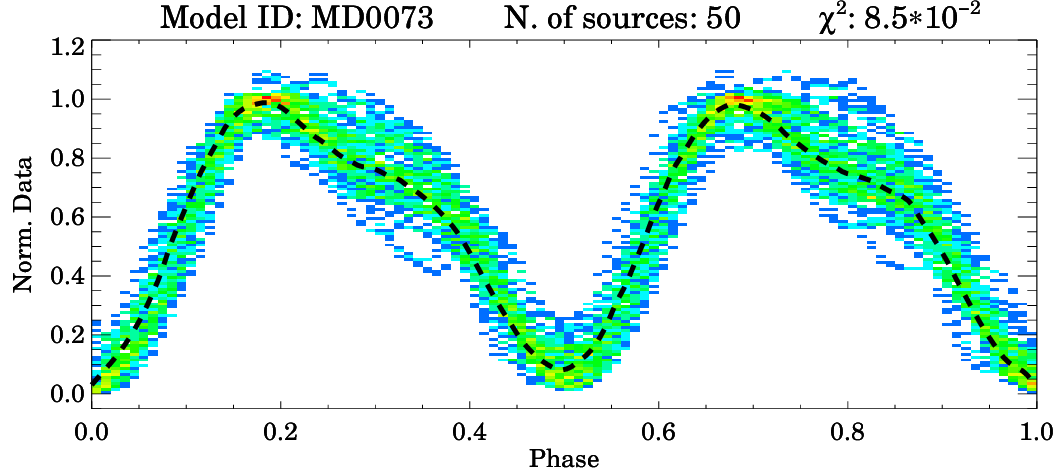}
  
  \includegraphics[width=0.33\textwidth]{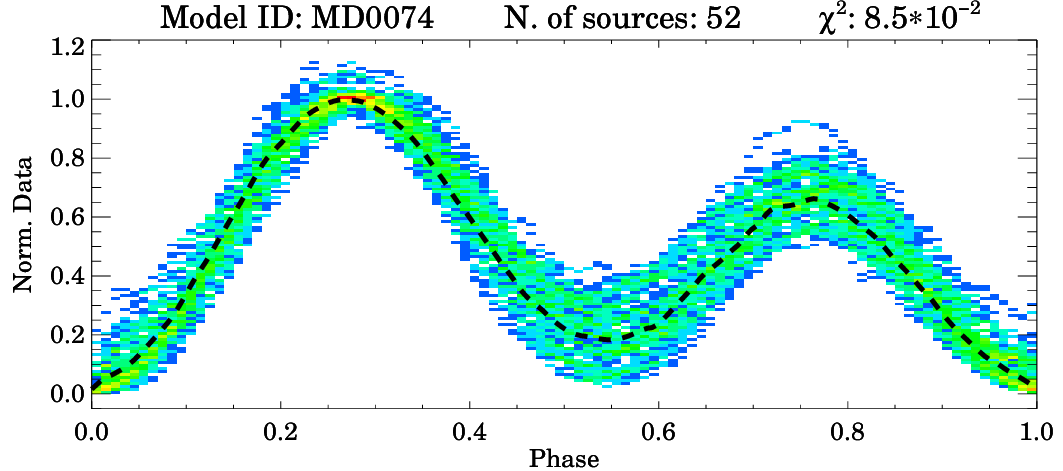}
  \includegraphics[width=0.33\textwidth]{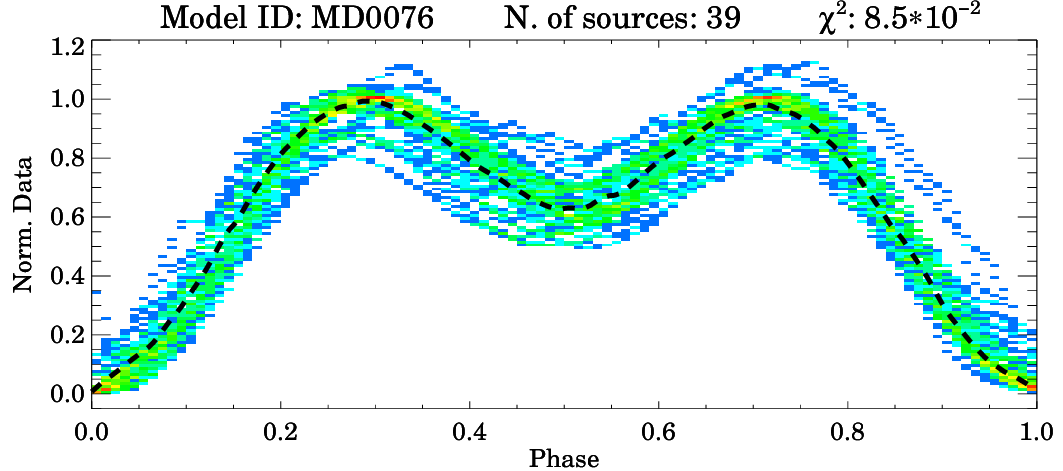}
  \includegraphics[width=0.33\textwidth]{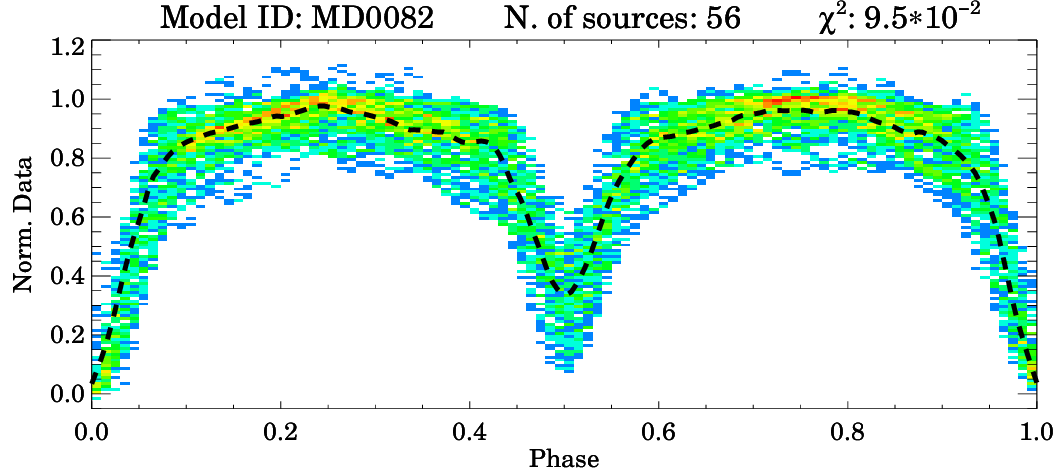}
  
  \includegraphics[width=0.33\textwidth]{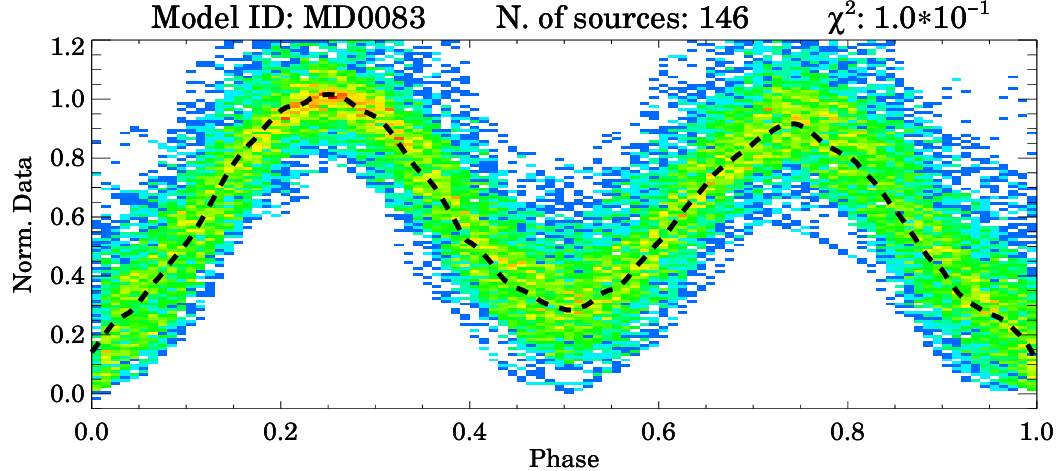}
  \includegraphics[width=0.33\textwidth]{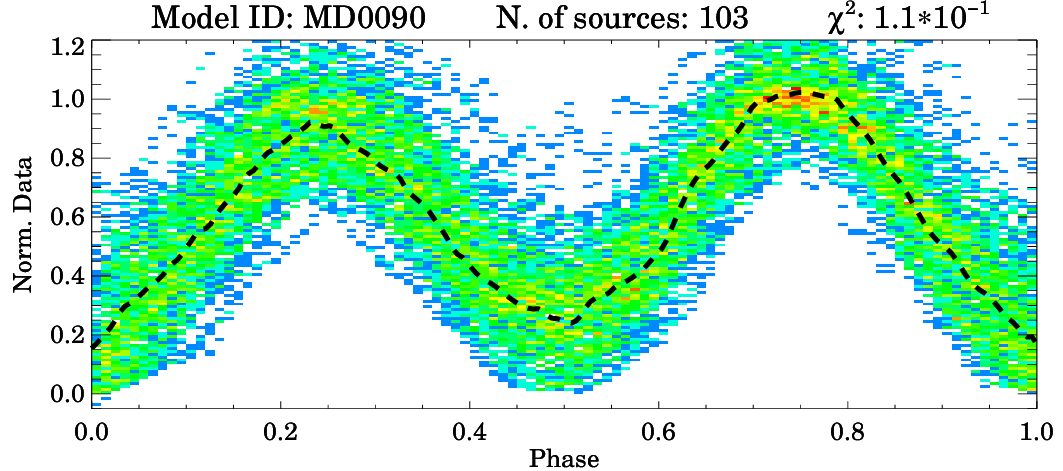}
  \includegraphics[width=0.33\textwidth]{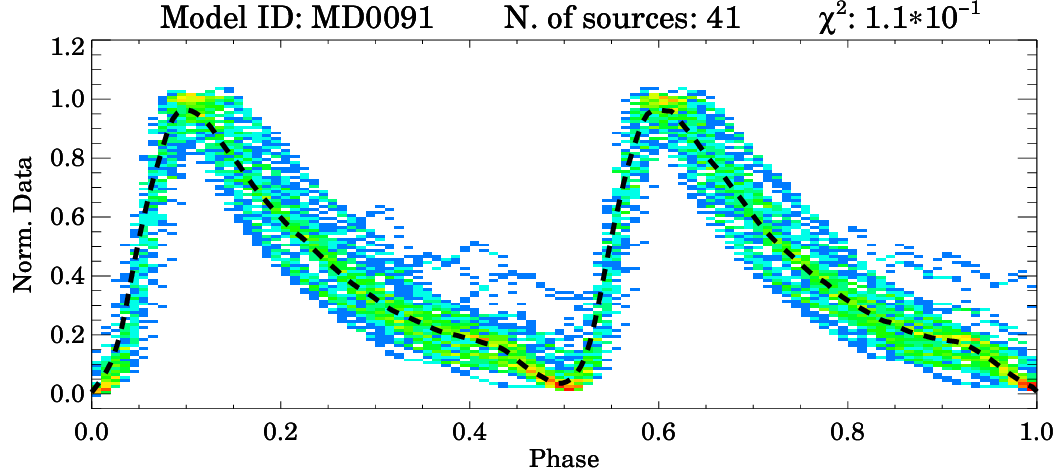}

  \includegraphics[width=0.33\textwidth]{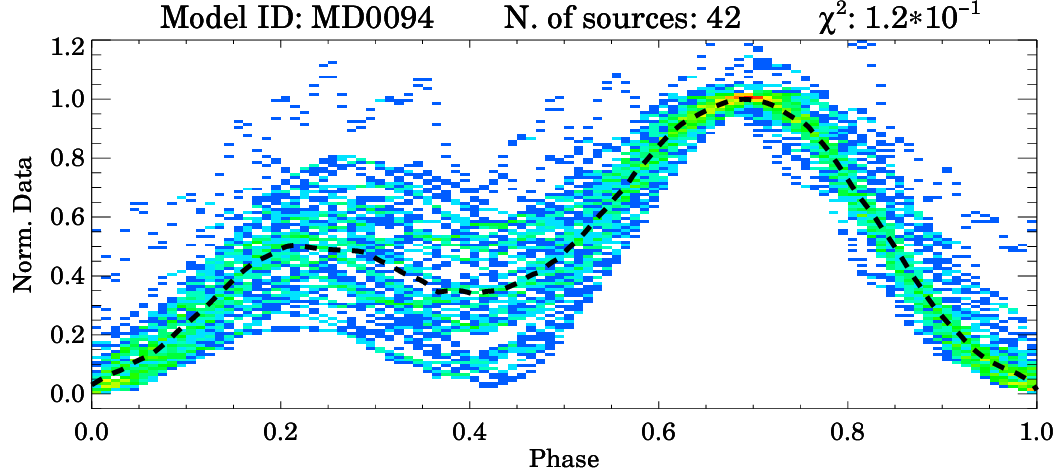}
  \includegraphics[width=0.33\textwidth]{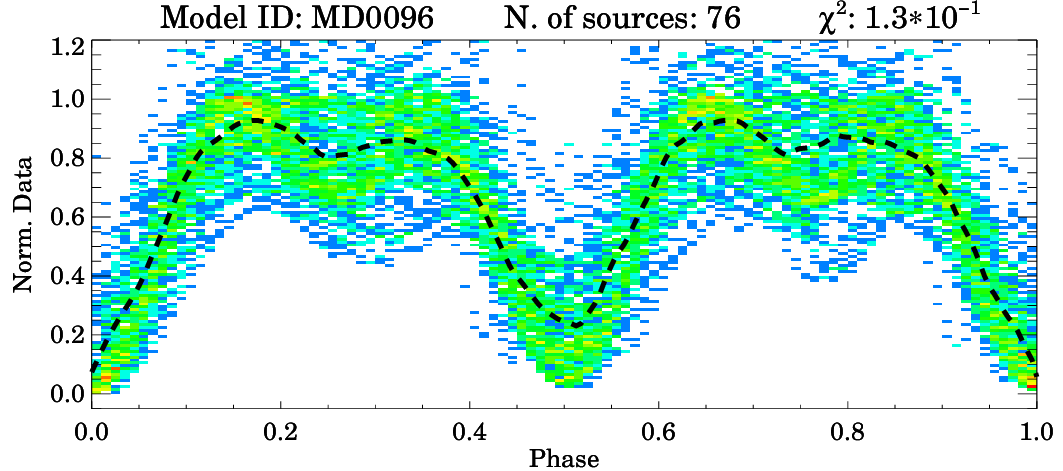}
  \includegraphics[width=0.33\textwidth]{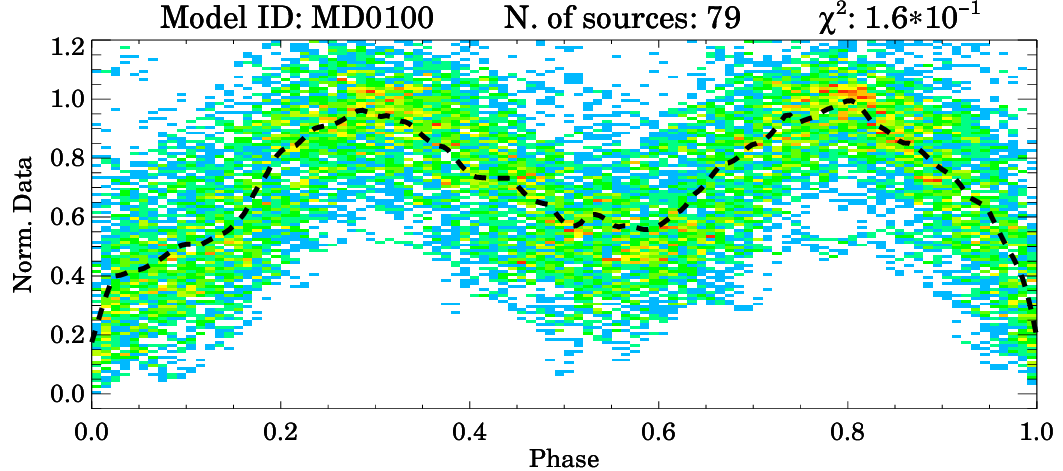}

    \caption{CoRoT phase diagrams used for classifying the \textit{CoRoT-CVSP}. The color density represents all sources included in the analysis. The black dashed line indicates the template model (see Section~\ref{sec_cross} for details). The model ID, number of sources, and maximum chi-square value used for selecting sources during model construction are indicated at the top of each panel. All template model data can be accessed through the CDS.}
  \label{fig_ModelCorot}
\end{figure*}

\begin{table*}
\caption{The first ten entries of the \textit{CoRoT-CVSP} catalog. }
\centering
\resizebox{\textwidth}{!}{ % Adjusts table to fit the page width
\begin{tabular}{l c c c c c c c c c c c c c c c c}
\hline
CoRoT ID & CoRoT Run & Ra (J2000) &  Dec (J2000) & SP & LC & Period (d) &  A (mmg) & TYPE & Nm & CoRoT mag & G$_{mag}$ & T$_{mag}$ & T$_{\rm eff}$ & log g & Dist & C/A   \\
\hline
631731296 & LRc07 & 278.69392 & 7.0241180 & K2 & III & 0.390881 & 4.111 & EW & 1 & 12.88 & 12.55 & 11.58 & 4087 & -999 & 2651 & c \\
631731518 & LRc07 & 278.73142 & 7.0092740 & F8 & IV & 0.048532 & 4.157 & DSCT & 1 & 14.13 & 13.89 & 13.41 & 6438 & 4.012 & 1204 & c \\
631899871 & LRc07 & 277.71636 & 7.1552230 & G0 & III & 2.116812 & 5.646 & EA & * & 12.06 & 12.01 & 11.45 & 5180 & 4.621 & 166.535 & c \\
631900682 & LRc07 & 277.84683 & 7.1625390 & A7 & V & 0.034036 & 5.967 & DSCT & 1 & 13.21 & 12.94 & 12.68 & 7787 & 3.987 & 1471 & c \\
631901153 & LRc07 & 277.91420 & 7.1050230 & A5 & V & 0.083381 & 8.370 & DSCT & 1 & 14.19 & 13.98 & 13.65 & 7390 & 3.610 & 3071 & c \\
102582083 & LRa01 & 100.23349 & -1.5260100 & A5 & IV & 0.572381 & 7.749 & GDOR & 2 & 13.04 & 13.09 & 12.82 & 7993 & 4.095 & 1400 & a \\
632083664 & LRc07 & 277.70350 & 7.2951310 & F5 & V & 0.040866 & 7.048 & DSCT & 12 & 13.25 & 13.03 & 12.64 & 6734 & 3.823 & 1277 & c \\
632084666 & LRc07 & 277.86796 & 7.2259820 & F8 & IV & 0.943738 & 10.556 & GDOR & 1 & 13.79 & 13.61 & 13.19 & 6653 & 4.130 & 1080 & c \\
632279463 & LRc07 & 277.53904 & 7.3959660 & G5 & IV & 0.983028 & 4.505 & EA & 31 & 12.57 & 12.38 & 11.88 & 5964 & 3.731 & 736 & c \\
102582304 & LRa01 & 100.23499 & 0.0789240 & A5 & V & 0.373380 & 10.155 & GDOR & 1 & 14.93 & 15.03 & 14.49 & 6670 & 4.078 & 1779 & a \\
\hline
\hline
\end{tabular}
}
\tablefoot{The complete table is available at the CDS.}
\label{table_data}
\end{table*}

\begin{figure}
  \centering
  \includegraphics[width=0.48\textwidth]{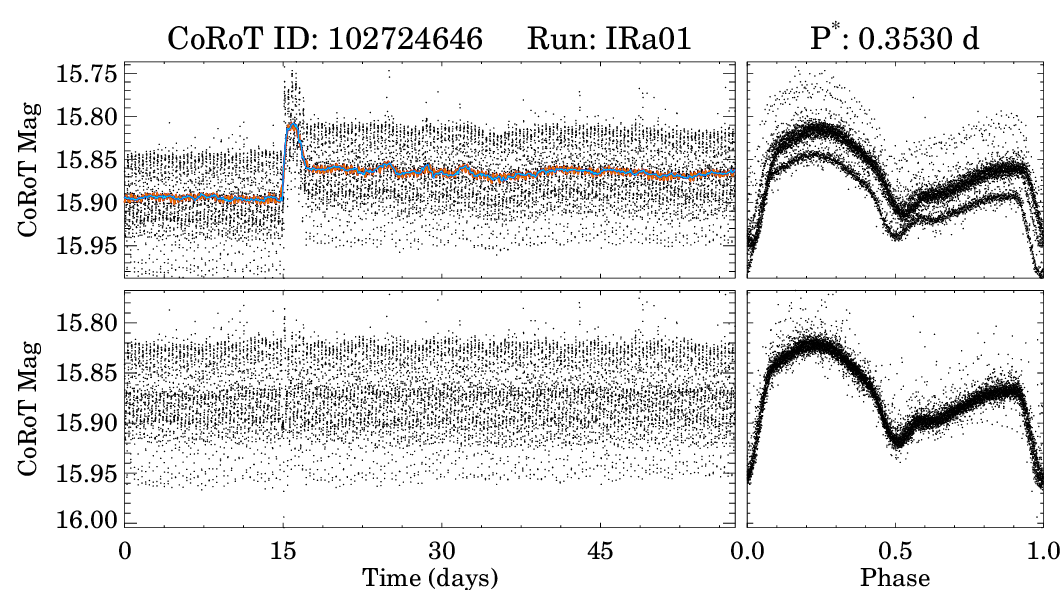}
  \includegraphics[width=0.48\textwidth]{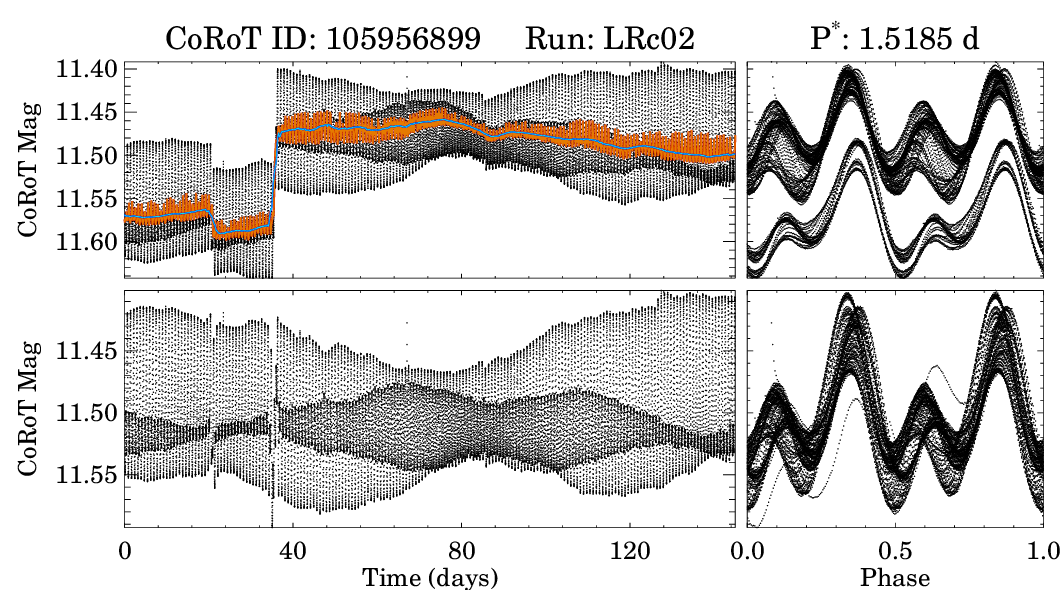}
  \caption{Example of CoRoT LCs (left panels) and phase diagram (right panels) of the "atypical" sources found in our sample.}
  \label{fig_CoRoTStrange}
\end{figure}

The LC-SSM approach enables an efficient and systematic grouping of similar light curve (LC) shapes, providing a time-effective alternative to the labor-intensive process of visual inspection \citep[][]{FerreiraLopes-2015wfcam,FerreiraLopes-2020-VIVA,Nikzat-2022}. For example, summarizing a sample of approximately 10,000 sources into 175, 105, 90, and 90 template models (for \(N_{S} = 20\), \(N_{S} = 40\), \(N_{S} = 60\), and \(N_{S} = 100\), respectively) required 10.5, 8.2, 7.0, and 5.0 hours. 
A minimum \( \chi^2 \) value of 0.01 was adopted in our analysis, representing a trade-off between the number of template models and maintaining sufficient descriptive accuracy, as illustrated in Fig.~4 (e.g., MD0001 vs. MD0100). The analysis was performed on a machine equipped with an Intel® Core™ i7-4510U CPU and 16~GB of RAM. Execution time could be further reduced with a more powerful machine or by optimizing the code. For this study, we selected \(N_{S} = 40\), resulting in a total of 102 template models. All template models are normalized to the [0, 1] interval, adopting the phase corresponding to the minimum magnitude as phase zero. The \(\chi^2\) values computed between each pair of models exceed 0.01, which is the threshold used to distinguish them. Considering MD0001 as the reference model, the \(\chi^2\) values relative to the other templates range from 0.018 to 0.38. This wide range highlights the diversity of light curve shapes represented by the template models.

The sources were grouped according to these template models, and a visual inspection of the LCs and their folded diagrams was conducted side by side, considering both raw and processed data (see Fig.~\ref{fig_CorotLCs}). This inspection process facilitated the efficient identification and classification of variable stars. The main conclusions are summarized below:

\begin{itemize}
    \item Approximately \( 9\% \) of the sources in our initial sample were removed during the visual inspection process. These sources exhibited issues such as residual bounces, trends, or other data artifacts that rendered their SNR values unreliable.
    
    \item The majority of the removed sources (\(\sim95\%\)) belonged to the short CoRoT runs, and \(\sim90\%\) had \( \mathrm{SNR} < 3 \). The shorter time coverage in these runs reduced the significance of detected signals, leading to a higher rate of misidentifications.
        
    \item All sources associated with models where \( \chi^2 < 1.0 \times 10^{-1} \) exhibited reliable signals (i.e., signals confirmed in the light curves and not affected by data issues). Higher \( \chi^2 \) values were typically found for signals affected by data artifacts, resulting in uncommon shapes. This outcome is expected, as the \( \chi^2 \) value quantifies the similarity between the signal and the associated model.
  
    \item Sources not assigned to any of the 102 template models were generally linked to data quality issues or unusual signal patterns. Approximately 2\% of these were due to incorrect periods or atypical signals. These cases may be of interest for follow-up, as they could represent rare phenomena such as double-mode pulsation or eclipsing binaries with strong asymmetries or reflection effects (see Fig.~\ref{fig_CoRoTStrange}).
    
    \item Folding the LC templates using twice the period computed by the LSG method (see Sect.~\ref{sec_dataset}) helped reconstruct the correct signal shape for eclipsing binaries, as the LSG method frequently identifies half the true period as dominant \citep[e.g.,][]{Papageorgiou-2018, FerreiraLopes-2021papIV, Christopoulou-2022}. This approach ensured proper classification of eclipsing binaries and preserved two cycles for other variable types. Template models such as MD0040, MD0043, MD0049, and MD0096 included these cases.
        
    \item Visual inspection of sources grouped by template models significantly streamlined and accelerated the classification process compared to inspecting all sources individually.
    
    \item Approximately 95\% of reliable signals could be identified without visual inspection by selecting only sources with \( \chi^2 \lesssim 1.5 \times 10^{-1} \).
\end{itemize}

These observations highlight the effectiveness of the LC-SSM approach. As a result, a final catalog of 9,272 stars was selected based on the criteria discussed above. This catalog, known as the \textit{CoRoT-CVSP} (CoRoT Catalog of Short-Period Variable Stars), provides valuable information on the identified variable stars. The main details of the catalog are presented in Table~\ref{table_data}, which can be accessed through the portal of Centre de Données astronomiques de Strasbourg (CDS)\footnote{\url{https://cdsweb.u-strasbg.fr/}}.
A useful parameter can be derived from the template model to determine whether the correct variability period is the true period or its harmonic. The \( \chi^{2}_{F} \) parameter is defined as the standard deviation of \( y_i - y_{i+N/2} \), where \( N \) is the number of components used in the templates model (\( N = 200 \) in our case). This parameter serves as a period flag by comparing the symmetry between the two halves of the phase diagram.

For single-period light curves, where both sides of the phase diagram are symmetric, \( \chi^{2}_{F} \simeq 0 \), as no asymmetry is present. However, for double-period light curves, where the two halves of the light curve differ, \( \chi^{2}_{F} \) increases with signal asymmetry. By applying a threshold of \( \chi^{2}_{F} > 5.0\times10^{-2} \), we identified 39 models for which the true variability period is likely twice the folded period.

To provide this information, we introduced a flag, denoted as \( F_{P} \), which classifies template model as either single-period (\( F_{P} = 1 \)) or double-period (\( F_{P} = 2 \)). However, for EW and ELL stars, the symmetry of their light curves may not be fully captured by the \( \chi^{2}_{F} \) value alone. Thus, relying solely on \( \chi^{2}_{F} \) may not always yield an accurate classification for these cases.

We found that most stars assigned to template models MD0013 and MD0036 belong to the EW type and were flagged as double-period (\( F_{P} = 2 \)). It is important to note that some template model, such as MD0046, may contain both EW and pulsating stars and were not flagged as \( F_{P} = 2 \). In total, 1,529 sources were flagged as \( F_{P} = 2 \), indicating that their period should be doubled.

We also investigated the sources located in the biased period regions (see Fig. \ref{fig_SNRPeriod}) with \(\rm Flag(Period) > 10\), which correspond to periods of approximately \(\sim1.40\) h, \(\sim0.85\) h, \(\sim0.57\) h, \(\sim24.07\) h, and their harmonics. The periods of \(\sim24\) h, \(\sim12\) h, and \(\sim6\) h were previously reported by \citet[][]{Degroote-2009} as being caused by instrumental drift. However, that study was based on a sample of 100 stars without variability signatures, whereas many sources in these period ranges exhibit clear variability signal with smooth phase diagrams. Approximately 96\% of the light curves in these regions come from only one LR field (out of 17) and seven SR fields (out of 8), with \(\sim42\%\) of them belonging to the LRa04 field. The remaining sources are distributed among SR fields as follows: SRa02 (\(\sim20\%\)), SRC02 (\(\sim12\%\)), SRa04 (\(\sim7\%\)), SRC03 (\(\sim6\%\)), SRa01 (\(\sim4\%\)), SRa03 (\(\sim3\%\)), and SRa01 (\(\sim2\%\)). Among these fields, only sources in LRa04 exhibit periods around \(\sim1.40\) h. We did not find any reported technical issues specifically related to these fields. The significant concentration of these periods suggests the presence of periodic noise associated with CCD temperature variations \citep[][]{Lapeyrere-2006,Auvergne-2009}.

\section{Classification and discussions}\label{sec_ClassificationTest}	

To incorporate both previous and newly derived classifications, we adopted a multi-step approach. The LC-SSM method provides an initial estimate for various types of variable stars, serving as a helpful starting point. However, since it relies solely on the light curve morphology, it cannot reliably classify all types of variability. As a first step, we used classifications available in the \textit{AAVSO} and \textsc{SIMBAD} databases to construct a training set for our classification procedure (see Sect. \ref{sec_cross}). Subsequently, we compared and combined the preliminary classifications obtained from LC-SSM with those from the literature to derive a more robust and consistent final classification for each object (see Sects. \ref{sec_debosher}) and \ref{sec_Final_Classification}).

\subsection{Cross-matching and preliminary classification}\label{sec_cross}	

Approximately 1,739 \textit{CoRoT-CVSP} sources are listed in the \textit{AAVSO} International Variable Star Index \citep[VSX;][]{Watson-2014}, 1,961 in the \textsc{SIMBAD} database\footnote{Accessed in February 2025.} \citep[][]{simbad-2000}, and 2,114 in the catalog compiled by \citet[][]{Gavras-2023} (hereafter referred to as GavrasCatalog) having variability classification. The matches found in VSX, SIMBAD, and GavrasCatalog were combined to create a master catalog set with 2,324 known variable star types. Indeed, the LS-SSM method provided initial classifications for various types of variable stars, serving as supplementary information for a machine learning-based classification (MCL). To visualize the data, a t-Distributed Stochastic Neighbor Embedding \citep[t-SNE;][]{vanDerMaaten2008} approach was implemented using Scikit-learn \citep{Pedregosa2011scikit-learn}, with a perplexity parameter set to 45 and 5,000 iterations.

The projection was performed onto a lower 3-dimensional manifold, enabling the exploration of interesting clusters, such as variable stars with asymmetric light curves, total eclipsing binaries, and contact eclipsing binaries with spurious double periods. To specifically identify Beta Lyrae eclipsing binaries (EB) among known eclipsing binaries, the Modified Local Linear Embedded method \citep[MLLE;][]{Zhang2006MLLEML} was applied for dimensionality reduction, projecting the data onto a 3-dimensional manifold. Variable stars positioned close to each other in this space are likely to belong to the same class. The number of nearest neighbors was set to 10. After visually inspecting the projected light curve data, new labels were assigned to the best subtypes of eclipsing binaries (EW, EA, EB).

\begin{table}
\caption{The distribution of variable stars in the different classes given in the CoRoT-CVSP catalog. The variable star class follows the \textsc{simbad}'s variable list, distinguishing between previously known (N.P.K - based on cross-matches with VSX and simbad) and previously unknown (N.P.U) stars and sky position, specifically towards the Galactic center (N.C.) and anti-center (N.A.).}
\centering
\begin{tabular}{ l c c | c c}
\hline
Type &   N.P.K &  N.P.U & N.C. & N.A.    \\
\hline
BCEP & $83$ & $226$ & $108$ & $201$\\
$\delta$ Sct & $561$ & $2644$ & $1154$ & $2051$\\
$\delta$ Sct/EW & $24$ & $26$ & $19$ & $31$\\
$\delta$ Sct/GDOR & $47$ & $208$ & $76$ & $179$\\
EA & $471$ & $128$ & $279$ & $320$\\
EB & $476$ & $368$ & $437$ & $407$\\
EW & $231$ & $266$ & $329$ & $168$\\
EW/GDOR & $14$ & $29$ & $12$ & $31$\\
GDOR & $274$ & $1169$ & $485$ & $958$\\
OrionV & $10$ & $0$ & $0$ & $10$\\
ROT & $11$ & $8$ & $5$ & $14$\\
RRab & $27$ & $29$ & $48$ & $8$\\
RRc & $5$ & $1$ & $4$ & $2$\\
RRd & $1$ & $0$ & $1$ & $0$\\
SPB & $119$ & $248$ & $69$ & $298$\\
TTau & $32$ & $0$ & $0$ & $32$\\
other & $236$ & $1302$ & $696$ & $840$\\
Total   &  $2622$  &  $6650$  &  $3722$  &  $5550$ \\ 
\hline\hline
\end{tabular} 
\label{table_classification}
\end{table}

A semi-supervised machine learning approach was applied using the Label Spreading method \citep[][]{Zhou2003LearningWL}. The known sample was divided into labeled and unlabeled sets, with 20\% of the data designated as a test set. The LS algorithm iteratively propagated label information from labeled points to their neighbors until convergence was reached or the maximum number of iterations was completed. The final classifications for the unlabeled points were determined based on the information obtained at the end of the iterative process. For the Scikit-learn implementation of the LS algorithm, the number of nearest neighbors was set to 10, and the gamma parameter was set to 2000.

We classified the variable stars into the following types and subtypes: EW, EA, EB, $\delta$ Scuti (DSCT), $\gamma$ Doradus (GDOR), and RR Lyrae (RRab, RRc, and RRd) following the matched data with Simbad and VSX datasets. Stars with ambiguous variability signatures were classified as \textit{other}. The accuracy of the classification was assessed using 5-fold cross-validation on the labeled subset, which was derived from cross-matched sources in VSX and SIMBAD. Performance metrics, including precision, recall, and F1-score, were computed to evaluate the model.  The overall classification accuracy was approximately 83\%, with the highest accuracy for EA-type stars (\(\sim93\%\)) and the lowest for $\gamma$ Doradus and RR Lyrae stars, at \(\sim73\%\) and \(\sim70\%\), respectively. 

Table \ref{table_classification} presents an overview of our final classification of \textit{CoRoT-CVSP} variable stars, including the number of both previously known and newly detected sources (see Sect. \ref{sec_Final_Classification}). It is important to note that \textit{SIMBAD} and \textit{VSX} compile most variable star detection catalogs, meaning some sources identified as "new" in this study may have prior identifications in the literature. Additionally, sources with limited classification information, such as those labeled as "Star" or "Var", were categorized as "unknown" sources.  

A note of caution is warranted regarding the limitations of the MAM used in this study. The MAM tends to remove signals with periods greater than approximately 2 days, as illustrated in Fig. \ref{fig_CorotLCs}. Therefore, our classification should be interpreted as relating to the short-period signals present in the light curves. In cases where a signal contains both a short period (\( P < 1 \) day) and a long period (\( P > 1 \) day), it is expected that rankings from past and present studies would align when derived from light curves corresponding to the same period. This consideration is crucial when comparing our classification results with those of other studies and in interpreting the nature of the identified variable stars.

\subsection{Benchmarking against previous classifications}\label{sec_debosher}	

The classification scheme developed by \citet[][]{Debosscher-2007,Debosscher-2009} (DCL)  was employed both as a benchmark for evaluating our classification and as a means to refine previous categorizations. Their approach relies on period analysis and harmonic fitting, using a training set derived from OGLE data \citep[][]{Walkowicz-2013}, which encompasses a broad range of variable star types, including BCEP ($\beta$-Cephei), BE (Be stars), CLCEP (Classical Cepheids), $\delta$ Sct, ECL (eclipsing binaries, all types), ELL (ellipsoidal variables), GDOR, RRab, RRc, RRd, and SPB (Slowly Pulsating B stars), among others. Indeed, DCL groups all eclipsing binaries into a single class (ECL), whereas our LC-SSM method distinguishes between the three primary types of eclipsing binaries: EA (Algol-type), EB ($\beta$ Lyrae-type), and EW (W Ursae Majoris-type). This distinction allows for a more nuanced analysis of the variability properties across different binary morphologies.

The DCL method employs a second-order polynomial fit, which is insufficient for addressing specific data quality issues. In cases where the data are relatively clean and the detected frequencies are consistent, both the MCL and DCL methods are expected to produce similar classification results. However, incorporating a probability factor significantly improves the reliability of the classifications. For instance, stars identified as $\delta$ Sct variables with a probability greater than 0.8 are considered robust classifications, as noted by \citet[][]{Michel-2017}. Considering these constraints, we selected all variable stars classified with a probability greater than $0.8$ by the DCL method and whose dominant frequencies were also recovered by the MCL approach. This subset was used to compare the two classification methods. The similarity of periods obtained by the DCL and MCL methods in this study suggests that the light curves have minor issues, and the results from both methods should generally be in agreement. Comparing the outcomes of the DCL and MCL methods provides insight into their effectiveness in identifying different classes of variable stars. Conducting these methods independently and using different parameters helps mitigate potential biases, ensuring a more robust evaluation of their performance. 

Table \ref{table_DebMcl} presents the classification results of the DCL and MCL methods for various types of variable stars. The concordance rates for GDOR and $\delta$ Sct stars represent the percentage of stars classified as such by both methods, which are approximately 88\% and 82\%, respectively. The mixed rate indicates the percentage of stars misclassified by both methods as the other type, at approximately 1.4\% and 16.4\% for GDOR and $\delta$ Sct stars, respectively. These findings suggest that both methods are relatively effective in identifying GDOR and $\delta$ Sct stars. However, achieving a complete separation of these classes requires further analysis, as they overlap in the parameter space concerning signal shape, period, and amplitude. Additional investigation, including other characteristics, may be necessary to achieve a more precise classification of these variable star types.  

Eclipsing binaries demonstrate a concordance rate greater than 77\%, indicating a relatively high level of agreement between the DCL and MCL methods in identifying this category. The highest concordance rate, around 97\%, is observed for EA-type stars. Conversely, the elevated mixing rate for GDOR stars can be attributed to several factors, including data issues and the similarity of GDOR light curves to those of EA/EB/EW stars.

\begin{table}
\caption{Concordance rates from comparing the DCL method and our classification.}
\centering
\begin{tabular}{ l  c c c c c c}
\hline
MCL/DCL& $\delta$ Sct & ECL & ELL & GDOR & RRab & RRc/d   \\
\hline
$\delta$ Sct & $82.3$ & $0.9$ & $0.4$ & $16.4$ & $0.0$ & $0.1$  \\
EA & $0.0$ & $97.8$ & $0.0$ & $2.2$ & $0.0$ & $0.0$  \\
EB & $13.4$ & $76.7$ & $0.6$ & $9.3$ & $0.0$ & $0.0$  \\
EW & $2.6$ & $79.7$ & $3.2$ & $10.0$ & $0.0$ & $4.5$  \\
GDOR & $1.4$ & $7.0$ & $3.1$ & $88.0$ & $0.0$ & $0.6$ \\
RRab & $0.0$ & $0.0$ & $11.1$ & $11.1$ & $77.8$ & $0.0$ \\
other & $8.3$ & $53.0$ & $3.2$ & $34.6$ & $0.0$ & $0.9$ \\
\hline\hline
\end{tabular} 
\label{table_DebMcl}
\end{table}

\subsection{Refined classification results}\label{sec_Final_Classification}	

By integrating the results from the DCL, MCL, and LC-SSM methods, we obtained a more comprehensive and robust classification of the \textit{CoRoT-CVSP} variable stars. The final classification scheme was based on the following criteria:

\begin{itemize}
    \item Only variable stars classified with a probability greater than 0.8 by the DCL method and with matching frequencies identified by the MCL method were considered. This subset represents approximately 85\% of the total number of variables analyzed by the DCL method, ensuring higher classification reliability and minimizing ambiguities.    
    \item Classifications from the DCL and MCL methods were merged to form a unified label. In cases of disagreement between the two methods, both classifications were retained.
    \item A classification of \textit{other} was assigned in cases where both the DCL and MCL methods identified the star as MISC or \textit{other}.  
    \item After merging the results from the DCL and MCL methods, the sources were grouped according to their variability class and associated with LC-SSM shape models. Stars with clear signatures of eclipsing binaries or RRab were re-labeled, whereas stars with ambiguous morphologies remained labeled as \textit{other}.
    \item Finally, for stars labeled as \textit{MISC} or \textit{other} after the previous steps, we adopted the classification available in external catalogs such as \textsc{SIMBAD}, VSX, or the GavrasCatalog, if available.
\end{itemize}

This multi-method approach allowed us to leverage the strengths of each technique. The LC-SSM method proved to be the most efficient, capturing nearly 98\% of the light curve shapes found in the \textit{CoRoT-CVSP} dataset. Its use significantly streamlined the classification process, reducing the need for visual inspection to only 102 representative light curves. It was particularly effective in identifying clear signals such as eclipsing binaries and RRab stars.

However, for some cases with ambiguous or overlapping features, additional analysis was required to improve classification reliability. In contrast, the MCL method served as a valuable tool for comparison and helped validate classifications suggested by the LC-SSM and DCL approaches. Although the MLC approach may be incomplete if the training set lacks certain variable star types, it still provides valuable classifications for further analysis and comparison with other methods.  

As mentioned earlier, data issues can affect classification accuracy. In particular, GDOR light curves may mimic those of eclipsing binaries, especially when the GDOR signal has low SNR or when the distinction between the primary and secondary eclipses is unclear. Visual inspection of light curves—such as those shown in Fig.~\ref{fig_ModelCorot} for M0096 and M0057—can reveal subtle features that help distinguish GDOR stars from eclipsing binaries. Similar challenges arise in the classification of RRab stars, where overlapping characteristics may lead to misidentification. Therefore, visual inspection remains essential for resolving ambiguities and improving classification reliability.

\section{Results}\label{sec_results}	

The \textit{CoRoT-CVSP} catalog, the main product of this study, contains a total of 9,272 variable stars, of which 6,249 are not listed in the \textsc{VSX}, \textsc{SIMBAD}, or GavrasCatalog databases. The catalog also includes atmospheric parameters from \textit{TESS}\footnote{Transiting Exoplanet Survey Satellite} \citep[][]{Ricker-2015} and \textit{Gaia} \citep[][]{Gaia-2016,Gaia-2021} DR3\footnote{Data Release 3}. These additional parameters offer important information about the physical properties and variability behavior of the stars, particularly those located in opposite directions of the Milky Way.

Our analysis and classification procedure offer a homogeneous and reliable assessment of these variable stars, addressing the primary data challenges associated with CoRoT observations for variable stars with periods shorter than one day. The resulting catalog represents a valuable resource for studying various variable star classes, including RR Lyrae stars, eclipsing binaries, and pulsating stars, among others. The sample consists of 5,550 stars located toward the Galactic center and 3,722 stars toward the Galactic anti-center (see Table~\ref{table_classification}). To facilitate further analysis, we define two subsamples: \textit{CoRoT-CVSP-C} for stars in the Galactic center region and \textit{CoRoT-CVSP-A} for those in the Galactic anti-center.

\begin{figure}
  \centering
  \includegraphics[width=0.47\textwidth]{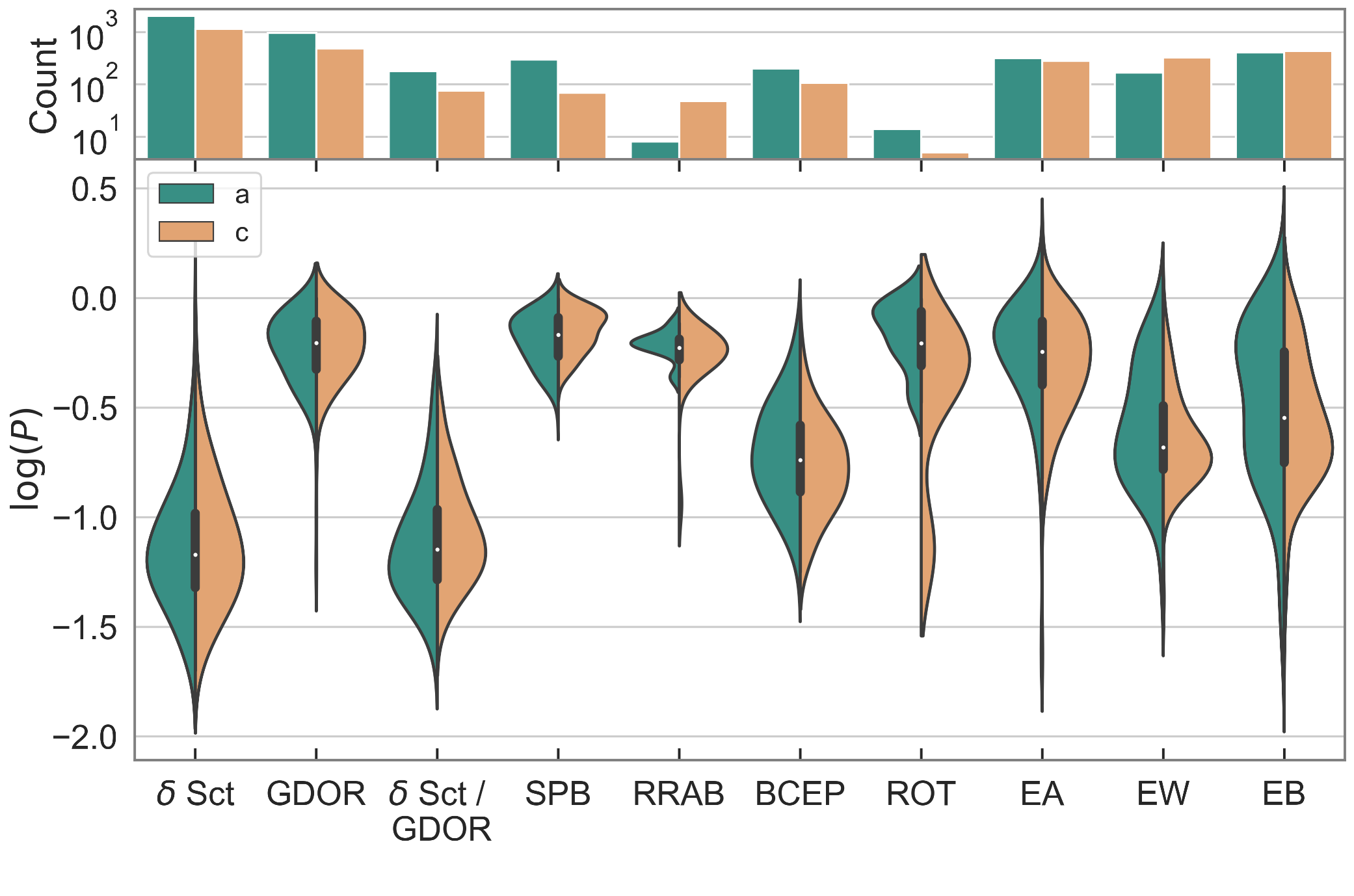}
  \includegraphics[width=0.47\textwidth]{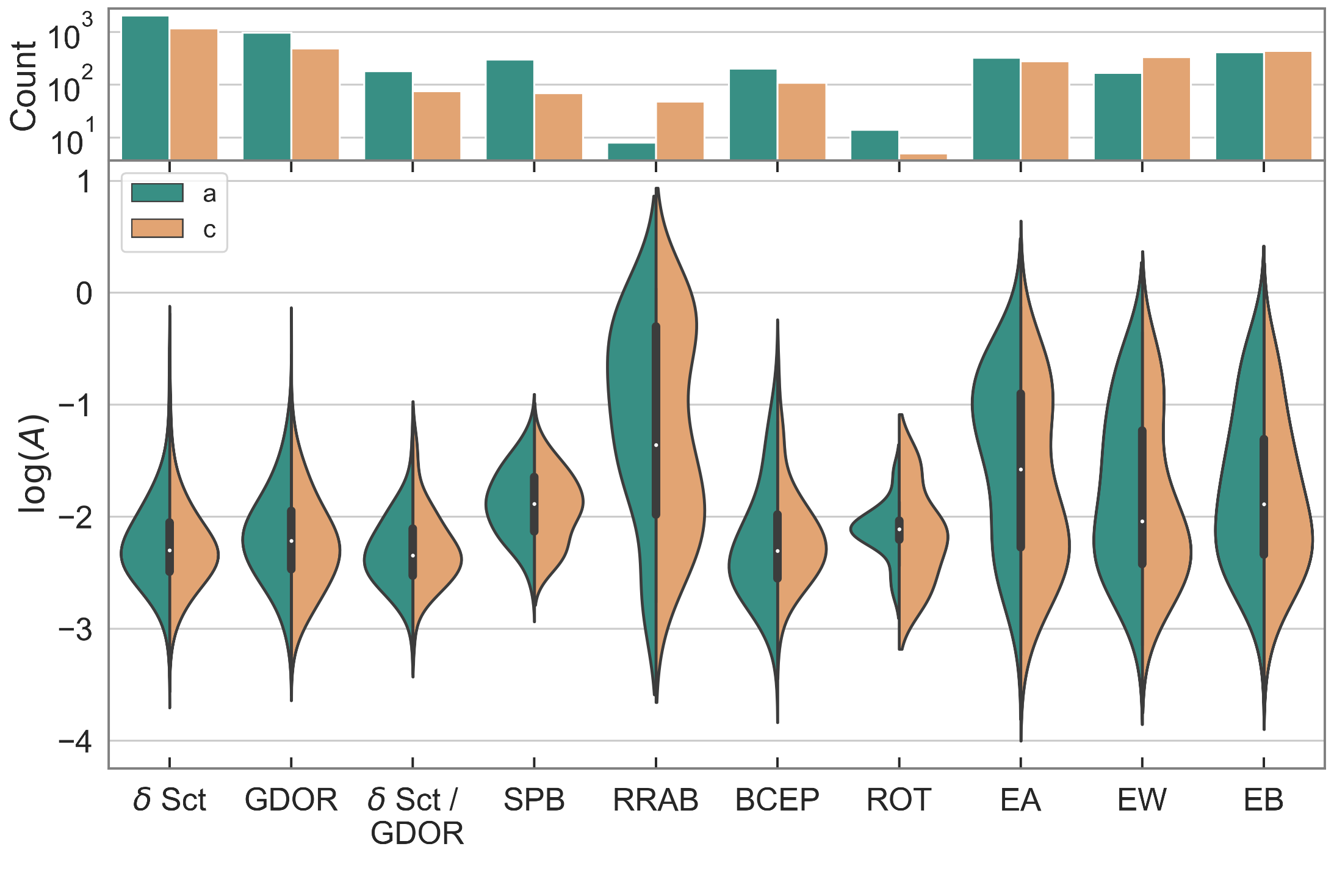}
  \caption{Violin plot  illustrating the distribution of period and amplitude for the \textit{CoRoT-CVSP-C} (in pale-orange) and \textit{CoRoT-CVSP-A} (in teal) samples.}
  \label{fig_histograms}
\end{figure}

Figure~\ref{fig_histograms} presents violin plots illustrating the distributions of period and amplitude for our sample of variable stars. The different colors represent the positions of these stars in the Galactic center (\textit{CoRoT-CVSP-C}) and anti-center (\textit{CoRoT-CVSP-A}). These violin plots reveal a significant disparity in the period and amplitude distributions between the \textit{CoRoT-CVSP-C} and \textit{CoRoT-CVSP-A} samples. Interestingly, variable star classifications often overlook these variations \citep[e.g.,][]{Marsakov-2011,DeMedeiros-2013,Zoccali-2017,Eilers-2022,Ratcliffe-2023}. This finding highlights the importance of integrating spatial information and other relevant parameters, such as metallicity and age, into the classification process. Accounting for Galactic location and additional factors, such as period and amplitude distributions, can lead to more refined and accurate classification models.

Comparing the \textit{CoRoT-CVSP-A} and \textit{CoRoT-CVSP-C} sub-samples provides crucial insights into the different variability characteristics of stars across Milky Way regions. To assess variations in period and amplitude, we applied the Kuiper test, also known as the invariant KS test \citep[e.g.,][]{Jetsu-1996,Paltani-2004}, to the two sub-samples. Notably, at an approximate distance of 4~kpc, a clear separation emerges between the center and anti-center \textit{CoRoT-CVSP} samples, according to Gaia distances. The KS test results for period and amplitude demonstrate a strong dependence on Galactic location, yielding \textit{p}-values smaller than \( \sim10^{-3} \) when comparing stars in the center and anti-center regions. These findings suggest a high probability that \textit{CoRoT-CVSP-A} and \textit{CoRoT-CVSP-C} originate from distinct stellar populations.

This result has broader implications for understanding the Milky Way, under the assumption that the regions analyzed are not exceptional. It aligns with the well-established variations in age and chemical composition across the Galaxy. The Galactic bulge, for instance, exhibits a wide range of metallicities, spanning from $-3.0$ to $+1.0$ dex \citep[e.g.,][]{Marsakov-2011,Do-2015,Zoccali-2017,Eilers-2022,Ratcliffe-2023}. Moreover, the anti-center region is expected to host older, more metal-poor stellar populations compared to the central region \citep[e.g.,][]{Feuillet-2019}. These complex spatial and elemental differences emphasize the necessity of considering diverse Galactic environments when analyzing stellar variability across the Milky Way.

\section{Conclusions}\label{sec_conclusions}
This study presents the results of a search for stars with variability periods shorter than one day, based on an extensive analysis of CoRoT mission data. A total of 9,272 variable stars were identified, including 6,249 not listed in SIMBAD or VSX, primarily classified through comparison with previously known variable types. Among them, we identified 309 $\beta$ Cephei, 3,105 $\delta$ Scuti, 599 Algol-type, 844 $\beta$ Lyrae, and 497 W Ursae Majoris eclipsing binaries, as well as 1,443 $\gamma$ Doradus, 63 RR Lyrae, and 32 T Tauri stars. The final sample, compiled into the \textit{CoRoT-CVSP} catalog, was built using a novel semi-supervised visual inspection method that achieved approximately 90\% efficiency in detecting variability signatures.

The \textit{CoRoT-CVSP} catalog, resulting from this study, is a valuable resource for studying stellar variability and the evolution of the Milky Way. It enables investigations of period-luminosity relations for $\delta$ Scuti, RR Lyrae, and $\gamma$ Doradus stars, and supports the analysis of reflection effects in eclipsing binaries with "atypical" light curve signatures.

This paper also presents a comprehensive classification of variable stars, combining multiple methods and literature-based information. The LC-SSM and MCL methods proved to be effective for summarizing and classifying large photometric datasets. With the integration of double-period analysis, the MCL approach enables the identification of variable stars through their smoother phase curves. These methodologies are broadly applicable beyond CoRoT, including to missions like Kepler and TESS. The LC-SSM approach is particularly suited for large-scale surveys where manual inspection is unfeasible.

In future papers of this series, we will extend our approach to identify CoRoT variable stars with periods longer than one day. This additional step will allow the detection of a broader range of variability types and contribute to a more comprehensive understanding of their properties.

\begin{acknowledgements}
C.E.F.L. and this project are supported by ANID’s Millennium Science Initiative through grants ICN12\_12009 and AIM23-0001, awarded to the Millennium Institute of Astrophysics (MAS); by ANID/FONDECYT Regular grant 1231637; by DIUDA 88231R11; by the LSST Discovery Alliance grant; and by GEMINI/ANID grant 32240028. D.H. acknowledges support from ANID through doctoral fellowship grant 21232262 for pursuing a Ph.D. M.C. acknowledges additional support from ANID's Basal project FB210003.  This study was partially funded by the Conselho Nacional de Desenvolvimento Cient\'ifico e Tecnol\'ogico (CNPq), the Coordena\c{c}\~ao de Aperfeiçoamento de Pessoal de N\'ivel Superior (CAPES)—Finance Code 001, and the CAPES-Print program. D.O.F. acknowledges CAPES graduate fellowships, and R.L.G. acknowledges a CNPq PDE fellowship. B.L.C.M. (grant No. 305804/2022-7), I.C.L. (grant No. 313103/2022-4), and J.R.M. (grant No. 308928/2019-9) acknowledge CNPq research fellowships. R.L.G. acknowledges CNPq postdoctoral fellowships (Grant Nos. 200031/2023-6 and 200744/2024-0).

\end{acknowledgements}

\bibliographystyle{aa}
\bibliography{MyLIB-AA.bib}

\begin{thebibliography}{73}
\expandafter\ifx\csname natexlab\endcsname\relax\def\natexlab#1{#1}\fi

\bibitem[{{Affer} {et~al.}(2012){Affer}, {Micela}, {Favata}, \&
  {Flaccomio}}]{Affer-2012}
{Affer}, L., {Micela}, G., {Favata}, F., \& {Flaccomio}, E. 2012, \mnras, 424,
  11

\bibitem[{{Alencar} {et~al.}(2010){Alencar}, {Teixeira}, {Guimar{\~a}es},
  {McGinnis}, {Gameiro}, {Bouvier}, {Aigrain}, {Flaccomio}, \&
  {Favata}}]{Alencar-2010}
{Alencar}, S.~H.~P., {Teixeira}, P.~S., {Guimar{\~a}es}, M.~M., {et~al.} 2010,
  \aap, 519, A88

\bibitem[{{Anders} {et~al.}(2017){Anders}, {Chiappini}, {Minchev}, {Miglio},
  {Montalb{\'a}n}, {Mosser}, {Rodrigues}, {Santiago}, {Baudin}, {Beers}, {da
  Costa}, {Garc{\'\i}a}, {Garc{\'\i}a-Hern{\'a}ndez}, {Holtzman}, {Maia},
  {Majewski}, {Mathur}, {Noels-Grotsch}, {Pan}, {Schneider}, {Schultheis},
  {Steinmetz}, {Valentini}, \& {Zamora}}]{Anders-2017}
{Anders}, F., {Chiappini}, C., {Minchev}, I., {et~al.} 2017, \aap, 600, A70

\bibitem[{{Anders} {et~al.}(2020){Anders}, {Minchev}, \&
  {Chiappini}}]{Anders-2020}
{Anders}, F., {Minchev}, I., \& {Chiappini}, C. 2020, in IAU General Assembly,
  257--257

\bibitem[{{Auvergne} {et~al.}(2009){Auvergne}, {Bodin}, {Boisnard}, {Buey},
  {Chaintreuil}, {Epstein}, {Jouret}, {Lam-Trong}, {Levacher}, {Magnan},
  {Perez}, {Plasson}, {Plesseria}, {Peter}, {Steller}, {Tiph{\`e}ne}, {Baglin},
  {Agogu{\'e}}, {Appourchaux}, {Barbet}, {Beaufort}, {Bellenger}, {Berlin},
  {Bernardi}, {Blouin}, {Boumier}, {Bonneau}, {Briet}, {Butler}, {Cautain},
  {Chiavassa}, {Costes}, {Cuvilho}, {Cunha-Parro}, {de Oliveira Fialho},
  {Decaudin}, {Defise}, {Djalal}, {Docclo}, {Drummond}, {Dupuis}, {Exil},
  {Faur{\'e}}, {Gaboriaud}, {Gamet}, {Gavalda}, {Grolleau}, {Gueguen},
  {Guivarc'h}, {Guterman}, {Hasiba}, {Huntzinger}, {Hustaix}, {Imbert},
  {Jeanville}, {Johlander}, {Jorda}, {Journoud}, {Karioty}, {Kerjean},
  {Lafond}, {Lapeyrere}, {Landiech}, {Larqu{\'e}}, {Laudet}, {Le Merrer},
  {Leporati}, {Leruyet}, {Levieuge}, {Llebaria}, {Martin}, {Mazy}, {Mesnager},
  {Michel}, {Moalic}, {Monjoin}, {Naudet}, {Neukirchner}, {Nguyen-Kim},
  {Ollivier}, {Orcesi}, {Ottacher}, {Oulali}, {Parisot}, {Perruchot},
  {Piacentino}, {Pinheiro da Silva}, {Platzer}, {Pontet}, {Pradines},
  {Quentin}, {Rohbeck}, {Rolland}, {Rollenhagen}, {Romagnan}, {Russ}, {Samadi},
  {Schmidt}, {Schwartz}, {Sebbag}, {Smit}, {Sunter}, {Tello}, {Toulouse},
  {Ulmer}, {Vandermarcq}, {Vergnault}, {Wallner}, {Waultier}, \&
  {Zanatta}}]{Auvergne-2009}
{Auvergne}, M., {Bodin}, P., {Boisnard}, L., {et~al.} 2009, \aap, 506, 411

\bibitem[{{Baeza-Villagra} {et~al.}(2025){Baeza-Villagra},
  {Rodr{\'\i}guez-Segovia}, {Catelan}, {Rest}, {Papageorgiou},
  {Mart{\'\i}nez-V{\'a}zquez}, {Valcarce}, {Ferreira Lopes}, \&
  {Bianco}}]{Baeza-Villagra-2025}
{Baeza-Villagra}, K., {Rodr{\'\i}guez-Segovia}, N., {Catelan}, M., {et~al.}
  2025, \aap, 694, A72

\bibitem[{{Baglin} {et~al.}(2007){Baglin}, {Auvergne}, {Barge}, {Michel},
  {Catala}, {Deleuil}, \& {Weiss}}]{Baglin-2007}
{Baglin}, A., {Auvergne}, M., {Barge}, P., {et~al.} 2007, in American Institute
  of Physics Conference Series, Vol. 895, Fifty Years of Romanian Astrophysics,
  ed. C.~{Dumitrache}, N.~A. {Popescu}, M.~D. {Suran}, \& V.~{Mioc}, 201--209

\bibitem[{{Benko}(2016)}]{Benko-2016}
{Benko}, J.~M. 2016, Information Bulletin on Variable Stars, 6189, 1

\bibitem[{{Bouchy} {et~al.}(2011){Bouchy}, {Deleuil}, {Guillot}, {Aigrain},
  {Carone}, {Cochran}, {Almenara}, {Alonso}, {Auvergne}, {Baglin}, {Barge},
  {Bonomo}, {Bord{\'e}}, {Csizmadia}, {de Bondt}, {Deeg}, {D{\'\i}az},
  {Dvorak}, {Endl}, {Erikson}, {Ferraz-Mello}, {Fridlund}, {Gandolfi},
  {Gazzano}, {Gibson}, {Gillon}, {Guenther}, {Hatzes}, {Havel}, {H{\'e}brard},
  {Jorda}, {L{\'e}ger}, {Lovis}, {Llebaria}, {Lammer}, {MacQueen}, {Mazeh},
  {Moutou}, {Ofir}, {Ollivier}, {Parviainen}, {P{\"a}tzold}, {Queloz}, {Rauer},
  {Rouan}, {Santerne}, {Schneider}, {Tingley}, \& {Wuchterl}}]{Bouchy-2011A}
{Bouchy}, F., {Deleuil}, M., {Guillot}, T., {et~al.} 2011, \aap, 525, A68

\bibitem[{{Carmo} {et~al.}(2020){Carmo}, {Ferreira Lopes}, {Papageorgiou},
  {Jablonski}, {Rodrigues}, {Drake}, {Cross}, \& {Catelan}}]{Carmo2020}
{Carmo}, A., {Ferreira Lopes}, C.~E., {Papageorgiou}, A., {et~al.} 2020,
  Boletin de la Asociacion Argentina de Astronomia La Plata Argentina, 61C, 88

\bibitem[{{Chadid} {et~al.}(2010){Chadid}, {Benk{\H{o}}}, {Szab{\'o}},
  {Papar{\'o}}, {Chapellier}, {Kolenberg}, {Poretti}, {Bono}, {Le Borgne},
  {Trinquet}, {Artemenko}, {Auvergne}, {Baglin}, {Debosscher}, {Grankin},
  {Guggenberger}, \& {Weiss}}]{Chadid-2010}
{Chadid}, M., {Benk{\H{o}}}, J.~M., {Szab{\'o}}, R., {et~al.} 2010, \aap, 510,
  A39

\bibitem[{{Chaintreuil} {et~al.}(2016){Chaintreuil}, {Bellucci}, {Baudin},
  {Ocvirk}, {Ballans}, {Landais}, {Ochsenbein}, {Orcesi}, \& {CoRot
  Team}}]{Chaintreuil-2016}
{Chaintreuil}, S., {Bellucci}, A., {Baudin}, F., {et~al.} 2016, {II.5 Where to
  find the CoRoT data?}, 109

\bibitem[{{Chiappini} {et~al.}(2015){Chiappini}, {Anders}, {Rodrigues},
  {Miglio}, {Montalb{\'a}n}, {Mosser}, {Girardi}, {Valentini}, {Noels},
  {Morel}, {Minchev}, {Steinmetz}, {Santiago}, {Schultheis}, {Martig}, {da
  Costa}, {Maia}, {Allende Prieto}, {de Assis Peralta}, {Hekker},
  {Theme{\ss}l}, {Kallinger}, {Garc{\'\i}a}, {Mathur}, {Baudin}, {Beers},
  {Cunha}, {Harding}, {Holtzman}, {Majewski}, {M{\'e}sz{\'a}ros}, {Nidever},
  {Pan}, {Schiavon}, {Shetrone}, {Schneider}, \& {Stassun}}]{Chiappini-2015}
{Chiappini}, C., {Anders}, F., {Rodrigues}, T.~S., {et~al.} 2015, \aap, 576,
  L12

\bibitem[{{Christopoulou} {et~al.}(2022){Christopoulou}, {Lalounta},
  {Papageorgiou}, {Ferreira Lopes}, {Catelan}, \& {Drake}}]{Christopoulou-2022}
{Christopoulou}, P.-E., {Lalounta}, E., {Papageorgiou}, A., {et~al.} 2022,
  \mnras, 512, 1244

\bibitem[{{Csizmadia} {et~al.}(2015){Csizmadia}, {Hatzes}, {Gandolfi},
  {Deleuil}, {Bouchy}, {Fridlund}, {Szabados}, {Parviainen}, {Cabrera},
  {Aigrain}, {Alonso}, {Almenara}, {Baglin}, {Bord{\'e}}, {Bonomo}, {Deeg},
  {D{\'\i}az}, {Erikson}, {Ferraz-Mello}, {Tadeu dos Santos}, {Guenther},
  {Guillot}, {Grziwa}, {H{\'e}brard}, {Klagyivik}, {Ollivier}, {P{\"a}tzold},
  {Rauer}, {Rouan}, {Santerne}, {Schneider}, {Mazeh}, {Wuchterl}, {Carpano}, \&
  {Ofir}}]{Csizmadia-2015}
{Csizmadia}, S., {Hatzes}, A., {Gandolfi}, D., {et~al.} 2015, \aap, 584, A13

\bibitem[{{Damiani} {et~al.}(2016){Damiani}, {Meunier}, {Moutou}, {Deleuil},
  {Ysard}, {Baudin}, \& {Deeg}}]{Damiani-2016}
{Damiani}, C., {Meunier}, J.~C., {Moutou}, C., {et~al.} 2016, \aap, 595, A95

\bibitem[{{De Medeiros} {et~al.}(2013){De Medeiros}, {Ferreira Lopes},
  {Le{\~a}o}, {Canto Martins}, {Catelan}, {Baglin}, {Vieira}, {Bravo},
  {Cort{\'e}s}, {de Freitas}, {Janot-Pacheco}, {Maciel}, {Melo}, {Osorio},
  {Porto de Mello}, \& {Valio}}]{DeMedeiros-2013}
{De Medeiros}, J.~R., {Ferreira Lopes}, C.~E., {Le{\~a}o}, I.~C., {et~al.}
  2013, \aap, 555, A63

\bibitem[{{Debosscher} {et~al.}(2007){Debosscher}, {Sarro}, {Aerts}, {Cuypers},
  {Vandenbussche}, {Garrido}, \& {Solano}}]{Debosscher-2007}
{Debosscher}, J., {Sarro}, L.~M., {Aerts}, C., {et~al.} 2007, \aap, 475, 1159

\bibitem[{{Debosscher} {et~al.}(2009){Debosscher}, {Sarro}, {L{\'o}pez},
  {Deleuil}, {Aerts}, {Auvergne}, {Baglin}, {Baudin}, {Chadid}, {Charpinet},
  {Cuypers}, {De Ridder}, {Garrido}, {Hubert}, {Janot-Pacheco}, {Jorda},
  {Kaiser}, {Kallinger}, {Kollath}, {Maceroni}, {Mathias}, {Michel}, {Moutou},
  {Neiner}, {Ollivier}, {Samadi}, {Solano}, {Surace}, {Vandenbussche}, \&
  {Weiss}}]{Debosscher-2009}
{Debosscher}, J., {Sarro}, L.~M., {L{\'o}pez}, M., {et~al.} 2009, \aap, 506,
  519

\bibitem[{{Degroote} {et~al.}(2009){Degroote}, {Aerts}, {Ollivier}, {Miglio},
  {Debosscher}, {Cuypers}, {Briquet}, {Montalb{\'a}n}, {Thoul}, {Noels}, {De
  Cat}, {Balaguer-N{\'u}{\~n}ez}, {Maceroni}, {Ribas}, {Auvergne}, {Baglin},
  {Deleuil}, {Weiss}, {Jorda}, {Baudin}, \& {Samadi}}]{Degroote-2009}
{Degroote}, P., {Aerts}, C., {Ollivier}, M., {et~al.} 2009, \aap, 506, 471

\bibitem[{{Deleuil} \& {Fridlund}(2018)}]{Deleuil-2018}
{Deleuil}, M. \& {Fridlund}, M. 2018, {CoRoT: The First Space-Based Transit
  Survey to Explore the Close-in Planet Population}, 79

\bibitem[{{Deleuil} {et~al.}(2009){Deleuil}, {Meunier}, {Moutou}, {Surace},
  {Deeg}, {Barbieri}, {Debosscher}, {Almenara}, {Agneray}, {Granet},
  {Guterman}, \& {Hodgkin}}]{Deleuil-2009}
{Deleuil}, M., {Meunier}, J.~C., {Moutou}, C., {et~al.} 2009, AJ, 138, 649

\bibitem[{{Do} {et~al.}(2015){Do}, {Kerzendorf}, {Winsor}, {St{\o}stad},
  {Morris}, {Lu}, \& {Ghez}}]{Do-2015}
{Do}, T., {Kerzendorf}, W., {Winsor}, N., {et~al.} 2015, \apj, 809, 143

\bibitem[{{Dolez} {et~al.}(2009){Dolez}, {Vauclair}, {Michel}, {Hui-Bon-Hoa},
  {Vauclair}, {Le Contel}, {Mathias}, {Poretti}, {Amado}, {Rainer}, {Samadi},
  {Baglin}, {Catala}, {Auvergne}, {Uytterhoeven}, \& {Valtier}}]{Dolez-2009}
{Dolez}, N., {Vauclair}, S., {Michel}, E., {et~al.} 2009, \aap, 506, 159

\bibitem[{{Eilers} {et~al.}(2022){Eilers}, {Hogg}, {Rix}, {Ness},
  {Price-Whelan}, {M{\'e}sz{\'a}ros}, \& {Nitschelm}}]{Eilers-2022}
{Eilers}, A.-C., {Hogg}, D.~W., {Rix}, H.-W., {et~al.} 2022, \apj, 928, 23

\bibitem[{{Ferreira Lopes} \& {Cross}(2016)}]{FerreiraLopes-2016papI}
{Ferreira Lopes}, C.~E. \& {Cross}, N.~J.~G. 2016, \aap, 586, A36

\bibitem[{{Ferreira Lopes} \& {Cross}(2017)}]{FerreiraLopes-2017papII}
{Ferreira Lopes}, C.~E. \& {Cross}, N.~J.~G. 2017, \aap, 604, A121

\bibitem[{{Ferreira Lopes} {et~al.}(2020){Ferreira Lopes}, {Cross}, {Catelan},
  {Minniti}, {Hempel}, {Lucas}, {Angeloni}, {Jablonsky}, {Braga}, {Le{\~a}o},
  {Herpich}, {Alonso-Garc{\'\i}a}, {Papageorgiou}, {Pichara}, {Saito},
  {Bradley}, {Beamin}, {Cort{\'e}s}, {De Medeiros}, \&
  {Russell}}]{FerreiraLopes-2020-VIVA}
{Ferreira Lopes}, C.~E., {Cross}, N.~J.~G., {Catelan}, M., {et~al.} 2020,
  \mnras, 496, 1730

\bibitem[{{Ferreira Lopes} {et~al.}(2018){Ferreira Lopes}, {Cross}, \&
  {Jablonski}}]{FerreiraLopes-2018papIII}
{Ferreira Lopes}, C.~E., {Cross}, N.~J.~G., \& {Jablonski}, F. 2018, \mnras,
  481, 3083

\bibitem[{{Ferreira Lopes} {et~al.}(2021){Ferreira Lopes}, {Cross}, \&
  {Jablonski}}]{FerreiraLopes-2021papIV}
{Ferreira Lopes}, C.~E., {Cross}, N.~J.~G., \& {Jablonski}, F. 2021, \mnras,
  501, 4123

\bibitem[{{Ferreira Lopes} {et~al.}(2015{\natexlab{a}}){Ferreira Lopes},
  {D{\'e}k{\'a}ny}, {Catelan}, {Cross}, {Angeloni}, {Le{\~a}o}, \& {De
  Medeiros}}]{FerreiraLopes-2015wfcam}
{Ferreira Lopes}, C.~E., {D{\'e}k{\'a}ny}, I., {Catelan}, M., {et~al.}
  2015{\natexlab{a}}, \aap, 573, A100

\bibitem[{{Ferreira Lopes} {et~al.}(2015{\natexlab{b}}){Ferreira Lopes},
  {Le{\~a}o}, {de Freitas}, {Canto Martins}, {Catelan}, \& {De
  Medeiros}}]{FerreiraLopes-2015cycles}
{Ferreira Lopes}, C.~E., {Le{\~a}o}, I.~C., {de Freitas}, D.~B., {et~al.}
  2015{\natexlab{b}}, \aap, 583, A134

\bibitem[{{Ferreira Lopes} {et~al.}(2015{\natexlab{c}}){Ferreira Lopes},
  {Neves}, {Le{\~a}o}, {de Freitas}, {Canto Martins}, {da Costa},
  {Paz-Chinch{\'o}n}, {Das Chagas}, {Baglin}, {Janot-Pacheco}, \& {De
  Medeiros}}]{FerreiraLopes-2015mgiant}
{Ferreira Lopes}, C.~E., {Neves}, V., {Le{\~a}o}, I.~C., {et~al.}
  2015{\natexlab{c}}, \aap, 583, A122

\bibitem[{{Feuillet} {et~al.}(2019){Feuillet}, {Frankel}, {Lind}, {Frinchaboy},
  {Garc{\'\i}a-Hern{\'a}ndez}, {Lane}, {Nitschelm}, \&
  {Roman-Lopes}}]{Feuillet-2019}
{Feuillet}, D.~K., {Frankel}, N., {Lind}, K., {et~al.} 2019, \mnras, 489, 1742

\bibitem[{{Gaia Collaboration} {et~al.}(2021){Gaia Collaboration}, {Brown},
  {Vallenari}, {Prusti}, {de Bruijne}, {Babusiaux}, {Biermann}, {Creevey},
  {Evans}, {Eyer}, {Hutton}, {Jansen}, {Jordi}, {Klioner}, {Lammers},
  {Lindegren}, {Luri}, {Mignard}, {Panem}, {Pourbaix}, {Randich}, {Sartoretti},
  {Soubiran}, {Walton}, {Arenou}, {Bailer-Jones}, {Bastian}, {Cropper},
  {Drimmel}, {Katz}, {Lattanzi}, {van Leeuwen}, {Bakker}, {Cacciari},
  {Casta{\~n}eda}, {De Angeli}, {Ducourant}, {Fabricius}, {Fouesneau},
  {Fr{\'e}mat}, {Guerra}, {Guerrier}, {Guiraud}, {Jean-Antoine Piccolo},
  {Masana}, {Messineo}, {Mowlavi}, {Nicolas}, {Nienartowicz}, {Pailler},
  {Panuzzo}, {Riclet}, {Roux}, {Seabroke}, {Sordo}, {Tanga}, {Th{\'e}venin},
  {Gracia-Abril}, {Portell}, {Teyssier}, {Altmann}, {Andrae}, {Bellas-Velidis},
  {Benson}, {Berthier}, {Blomme}, {Brugaletta}, {Burgess}, {Busso}, {Carry},
  {Cellino}, {Cheek}, {Clementini}, {Damerdji}, {Davidson}, {Delchambre},
  {Dell'Oro}, {Fern{\'a}ndez-Hern{\'a}ndez}, {Galluccio}, {Garc{\'\i}a-Lario},
  {Garcia-Reinaldos}, {Gonz{\'a}lez-N{\'u}{\~n}ez}, {Gosset}, {Haigron},
  {Halbwachs}, {Hambly}, {Harrison}, {Hatzidimitriou}, {Heiter},
  {Hern{\'a}ndez}, {Hestroffer}, {Hodgkin}, {Holl}, {Jan{\ss}en}, {Jevardat de
  Fombelle}, {Jordan}, {Krone-Martins}, {Lanzafame}, {L{\"o}ffler}, {Lorca},
  {Manteiga}, {Marchal}, {Marrese}, {Moitinho}, {Mora}, {Muinonen}, {Osborne},
  {Pancino}, {Pauwels}, {Petit}, {Recio-Blanco}, {Richards}, {Riello},
  {Rimoldini}, {Robin}, {Roegiers}, {Rybizki}, {Sarro}, {Siopis}, {Smith},
  {Sozzetti}, {Ulla}, {Utrilla}, {van Leeuwen}, {van Reeven}, {Abbas}, {Abreu
  Aramburu}, {Accart}, {Aerts}, {Aguado}, {Ajaj}, {Altavilla}, {{\'A}lvarez},
  {{\'A}lvarez Cid-Fuentes}, {Alves}, {Anderson}, {Anglada Varela}, {Antoja},
  {Audard}, {Baines}, {Baker}, {Balaguer-N{\'u}{\~n}ez}, {Balbinot}, {Balog},
  {Barache}, {Barbato}, {Barros}, {Barstow}, {Bartolom{\'e}}, {Bassilana},
  {Bauchet}, {Baudesson-Stella}, {Becciani}, {Bellazzini}, {Bernet}, {Bertone},
  {Bianchi}, {Blanco-Cuaresma}, {Boch}, {Bombrun}, {Bossini}, {Bouquillon},
  {Bragaglia}, {Bramante}, {Breedt}, {Bressan}, {Brouillet}, {Bucciarelli},
  {Burlacu}, {Busonero}, {Butkevich}, {Buzzi}, {Caffau}, {Cancelliere},
  {C{\'a}novas}, {Cantat-Gaudin}, {Carballo}, {Carlucci}, {Carnerero},
  {Carrasco}, {Casamiquela}, {Castellani}, {Castro-Ginard}, {Castro Sampol},
  {Chaoul}, {Charlot}, {Chemin}, {Chiavassa}, {Cioni}, {Comoretto}, {Cooper},
  {Cornez}, {Cowell}, {Crifo}, {Crosta}, {Crowley}, {Dafonte}, {Dapergolas},
  {David}, {David}, {de Laverny}, {De Luise}, {De March}, {De Ridder}, {de
  Souza}, {de Teodoro}, {de Torres}, {del Peloso}, {del Pozo}, {Delbo},
  {Delgado}, {Delgado}, {Delisle}, {Di Matteo}, {Diakite}, {Diener},
  {Distefano}, {Dolding}, {Eappachen}, {Edvardsson}, {Enke}, {Esquej}, {Fabre},
  {Fabrizio}, {Faigler}, {Fedorets}, {Fernique}, {Fienga}, {Figueras},
  {Fouron}, {Fragkoudi}, {Fraile}, {Franke}, {Gai}, {Garabato},
  {Garcia-Gutierrez}, {Garc{\'\i}a-Torres}, {Garofalo}, {Gavras}, {Gerlach},
  {Geyer}, {Giacobbe}, {Gilmore}, {Girona}, {Giuffrida}, {Gomel}, {Gomez},
  {Gonzalez-Santamaria}, {Gonz{\'a}lez-Vidal}, {Granvik},
  {Guti{\'e}rrez-S{\'a}nchez}, {Guy}, {Hauser}, {Haywood}, {Helmi}, {Hidalgo},
  {Hilger}, {H{\l}adczuk}, {Hobbs}, {Holland}, {Huckle}, {Jasniewicz},
  {Jonker}, {Juaristi Campillo}, {Julbe}, {Karbevska}, {Kervella}, {Khanna},
  {Kochoska}, {Kontizas}, {Kordopatis}, {Korn}, {Kostrzewa-Rutkowska},
  {Kruszy{\'n}ska}, {Lambert}, {Lanza}, {Lasne}, {Le Campion}, {Le Fustec},
  {Lebreton}, {Lebzelter}, {Leccia}, {Leclerc}, {Lecoeur-Taibi}, {Liao},
  {Licata}, {Lindstr{\o}m}, {Lister}, {Livanou}, {Lobel}, {Madrero Pardo},
  {Managau}, {Mann}, {Marchant}, {Marconi}, {Marcos Santos}, {Marinoni},
  {Marocco}, {Marshall}, {Martin Polo}, {Mart{\'\i}n-Fleitas}, {Masip},
  {Massari}, {Mastrobuono-Battisti}, {Mazeh}, {McMillan}, {Messina},
  {Michalik}, {Millar}, {Mints}, {Molina}, {Molinaro}, {Moln{\'a}r},
  {Montegriffo}, {Mor}, {Morbidelli}, {Morel}, {Morris}, {Mulone}, {Munoz},
  {Muraveva}, {Murphy}, {Musella}, {Noval}, {Ord{\'e}novic}, {Orr{\`u}},
  {Osinde}, {Pagani}, {Pagano}, {Palaversa}, {Palicio}, {Panahi}, {Pawlak},
  {Pe{\~n}alosa Esteller}, {Penttil{\"a}}, {Piersimoni}, {Pineau}, {Plachy},
  {Plum}, {Poggio}, {Poretti}, {Poujoulet}, {Pr{\v{s}}a}, {Pulone}, {Racero},
  {Ragaini}, {Rainer}, {Raiteri}, {Rambaux}, {Ramos}, {Ramos-Lerate}, {Re
  Fiorentin}, {Regibo}, {Reyl{\'e}}, {Ripepi}, {Riva}, {Rixon}, {Robichon},
  {Robin}, {Roelens}, {Rohrbasser}, {Romero-G{\'o}mez}, {Rowell}, {Royer},
  {Rybicki}, {Sadowski}, {Sagrist{\`a} Sell{\'e}s}, {Sahlmann}, {Salgado},
  {Salguero}, {Samaras}, {Sanchez Gimenez}, {Sanna}, {Santove{\~n}a},
  {Sarasso}, {Schultheis}, {Sciacca}, {Segol}, {Segovia}, {S{\'e}gransan},
  {Semeux}, {Shahaf}, {Siddiqui}, {Siebert}, {Siltala}, {Slezak}, {Smart},
  {Solano}, {Solitro}, {Souami}, {Souchay}, {Spagna}, {Spoto}, {Steele},
  {Steidelm{\"u}ller}, {Stephenson}, {S{\"u}veges}, {Szabados}, {Szegedi-Elek},
  {Taris}, {Tauran}, {Taylor}, {Teixeira}, {Thuillot}, {Tonello}, {Torra},
  {Torra}, {Turon}, {Unger}, {Vaillant}, {van Dillen}, {Vanel}, {Vecchiato},
  {Viala}, {Vicente}, {Voutsinas}, {Weiler}, {Wevers}, {Wyrzykowski}, {Yoldas},
  {Yvard}, {Zhao}, {Zorec}, {Zucker}, {Zurbach}, \& {Zwitter}}]{Gaia-2021}
{Gaia Collaboration}, {Brown}, A.~G.~A., {Vallenari}, A., {et~al.} 2021, \aap,
  649, A1

\bibitem[{{Gaia Collaboration} {et~al.}(2016){Gaia Collaboration}, {Prusti},
  {de Bruijne}, {Brown}, {Vallenari}, {Babusiaux}, {Bailer-Jones}, {Bastian},
  {Biermann}, {Evans}, {Eyer}, {Jansen}, {Jordi}, {Klioner}, {Lammers},
  {Lindegren}, {Luri}, {Mignard}, {Milligan}, {Panem}, {Poinsignon},
  {Pourbaix}, {Randich}, {Sarri}, {Sartoretti}, {Siddiqui}, {Soubiran},
  {Valette}, {van Leeuwen}, {Walton}, {Aerts}, {Arenou}, {Cropper}, {Drimmel},
  {H{\o}g}, {Katz}, {Lattanzi}, {O'Mullane}, {Grebel}, {Holland}, {Huc},
  {Passot}, {Bramante}, {Cacciari}, {Casta{\~n}eda}, {Chaoul}, {Cheek}, {De
  Angeli}, {Fabricius}, {Guerra}, {Hern{\'a}ndez}, {Jean-Antoine-Piccolo},
  {Masana}, {Messineo}, {Mowlavi}, {Nienartowicz}, {Ord{\'o}{\~n}ez-Blanco},
  {Panuzzo}, {Portell}, {Richards}, {Riello}, {Seabroke}, {Tanga},
  {Th{\'e}venin}, {Torra}, {Els}, {Gracia-Abril}, {Comoretto},
  {Garcia-Reinaldos}, {Lock}, {Mercier}, {Altmann}, {Andrae}, {Astraatmadja},
  {Bellas-Velidis}, {Benson}, {Berthier}, {Blomme}, {Busso}, {Carry},
  {Cellino}, {Clementini}, {Cowell}, {Creevey}, {Cuypers}, {Davidson}, {De
  Ridder}, {de Torres}, {Delchambre}, {Dell'Oro}, {Ducourant}, {Fr{\'e}mat},
  {Garc{\'\i}a-Torres}, {Gosset}, {Halbwachs}, {Hambly}, {Harrison}, {Hauser},
  {Hestroffer}, {Hodgkin}, {Huckle}, {Hutton}, {Jasniewicz}, {Jordan},
  {Kontizas}, {Korn}, {Lanzafame}, {Manteiga}, {Moitinho}, {Muinonen},
  {Osinde}, {Pancino}, {Pauwels}, {Petit}, {Recio-Blanco}, {Robin}, {Sarro},
  {Siopis}, {Smith}, {Smith}, {Sozzetti}, {Thuillot}, {van Reeven}, {Viala},
  {Abbas}, {Abreu Aramburu}, {Accart}, {Aguado}, {Allan}, {Allasia},
  {Altavilla}, {{\'A}lvarez}, {Alves}, {Anderson}, {Andrei}, {Anglada Varela},
  {Antiche}, {Antoja}, {Ant{\'o}n}, {Arcay}, {Atzei}, {Ayache}, {Bach},
  {Baker}, {Balaguer-N{\'u}{\~n}ez}, {Barache}, {Barata}, {Barbier}, {Barblan},
  {Baroni}, {Barrado y Navascu{\'e}s}, {Barros}, {Barstow}, {Becciani},
  {Bellazzini}, {Bellei}, {Bello Garc{\'\i}a}, {Belokurov}, {Bendjoya},
  {Berihuete}, {Bianchi}, {Bienaym{\'e}}, {Billebaud}, {Blagorodnova},
  {Blanco-Cuaresma}, {Boch}, {Bombrun}, {Borrachero}, {Bouquillon}, {Bourda},
  {Bouy}, {Bragaglia}, {Breddels}, {Brouillet}, {Br{\"u}semeister},
  {Bucciarelli}, {Budnik}, {Burgess}, {Burgon}, {Burlacu}, {Busonero}, {Buzzi},
  {Caffau}, {Cambras}, {Campbell}, {Cancelliere}, {Cantat-Gaudin}, {Carlucci},
  {Carrasco}, {Castellani}, {Charlot}, {Charnas}, {Charvet}, {Chassat},
  {Chiavassa}, {Clotet}, {Cocozza}, {Collins}, {Collins}, {Costigan}, {Crifo},
  {Cross}, {Crosta}, {Crowley}, {Dafonte}, {Damerdji}, {Dapergolas}, {David},
  {David}, {De Cat}, {de Felice}, {de Laverny}, {De Luise}, {De March}, {de
  Martino}, {de Souza}, {Debosscher}, {del Pozo}, {Delbo}, {Delgado},
  {Delgado}, {di Marco}, {Di Matteo}, {Diakite}, {Distefano}, {Dolding}, {Dos
  Anjos}, {Drazinos}, {Dur{\'a}n}, {Dzigan}, {Ecale}, {Edvardsson}, {Enke},
  {Erdmann}, {Escolar}, {Espina}, {Evans}, {Eynard Bontemps}, {Fabre},
  {Fabrizio}, {Faigler}, {Falc{\~a}o}, {Farr{\`a}s Casas}, {Faye}, {Federici},
  {Fedorets}, {Fern{\'a}ndez-Hern{\'a}ndez}, {Fernique}, {Fienga}, {Figueras},
  {Filippi}, {Findeisen}, {Fonti}, {Fouesneau}, {Fraile}, {Fraser}, {Fuchs},
  {Furnell}, {Gai}, {Galleti}, {Galluccio}, {Garabato}, {Garc{\'\i}a-Sedano},
  {Gar{\'e}}, {Garofalo}, {Garralda}, {Gavras}, {Gerssen}, {Geyer}, {Gilmore},
  {Girona}, {Giuffrida}, {Gomes}, {Gonz{\'a}lez-Marcos},
  {Gonz{\'a}lez-N{\'u}{\~n}ez}, {Gonz{\'a}lez-Vidal}, {Granvik}, {Guerrier},
  {Guillout}, {Guiraud}, {G{\'u}rpide}, {Guti{\'e}rrez-S{\'a}nchez}, {Guy},
  {Haigron}, {Hatzidimitriou}, {Haywood}, {Heiter}, {Helmi}, {Hobbs},
  {Hofmann}, {Holl}, {Holland}, {Hunt}, {Hypki}, {Icardi}, {Irwin}, {Jevardat
  de Fombelle}, {Jofr{\'e}}, {Jonker}, {Jorissen}, {Julbe}, {Karampelas},
  {Kochoska}, {Kohley}, {Kolenberg}, {Kontizas}, {Koposov}, {Kordopatis},
  {Koubsky}, {Kowalczyk}, {Krone-Martins}, {Kudryashova}, {Kull}, {Bachchan},
  {Lacoste-Seris}, {Lanza}, {Lavigne}, {Le Poncin-Lafitte}, {Lebreton},
  {Lebzelter}, {Leccia}, {Leclerc}, {Lecoeur-Taibi}, {Lemaitre}, {Lenhardt},
  {Leroux}, {Liao}, {Licata}, {Lindstr{\o}m}, {Lister}, {Livanou}, {Lobel},
  {L{\"o}ffler}, {L{\'o}pez}, {Lopez-Lozano}, {Lorenz}, {Loureiro},
  {MacDonald}, {Magalh{\~a}es Fernandes}, {Managau}, {Mann}, {Mantelet},
  {Marchal}, {Marchant}, {Marconi}, {Marie}, {Marinoni}, {Marrese},
  {Marschalk{\'o}}, {Marshall}, {Mart{\'\i}n-Fleitas}, {Martino}, {Mary},
  {Matijevi{\v{c}}}, {Mazeh}, {McMillan}, {Messina}, {Mestre}, {Michalik},
  {Millar}, {Miranda}, {Molina}, {Molinaro}, {Molinaro}, {Moln{\'a}r},
  {Moniez}, {Montegriffo}, {Monteiro}, {Mor}, {Mora}, {Morbidelli}, {Morel},
  {Morgenthaler}, {Morley}, {Morris}, {Mulone}, {Muraveva}, {Musella},
  {Narbonne}, {Nelemans}, {Nicastro}, {Noval}, {Ord{\'e}novic},
  {Ordieres-Mer{\'e}}, {Osborne}, {Pagani}, {Pagano}, {Pailler}, {Palacin},
  {Palaversa}, {Parsons}, {Paulsen}, {Pecoraro}, {Pedrosa}, {Pentik{\"a}inen},
  {Pereira}, {Pichon}, {Piersimoni}, {Pineau}, {Plachy}, {Plum}, {Poujoulet},
  {Pr{\v{s}}a}, {Pulone}, {Ragaini}, {Rago}, {Rambaux}, {Ramos-Lerate},
  {Ranalli}, {Rauw}, {Read}, {Regibo}, {Renk}, {Reyl{\'e}}, {Ribeiro},
  {Rimoldini}, {Ripepi}, {Riva}, {Rixon}, {Roelens}, {Romero-G{\'o}mez},
  {Rowell}, {Royer}, {Rudolph}, {Ruiz-Dern}, {Sadowski}, {Sagrist{\`a}
  Sell{\'e}s}, {Sahlmann}, {Salgado}, {Salguero}, {Sarasso}, {Savietto},
  {Schnorhk}, {Schultheis}, {Sciacca}, {Segol}, {Segovia}, {Segransan},
  {Serpell}, {Shih}, {Smareglia}, {Smart}, {Smith}, {Solano}, {Solitro},
  {Sordo}, {Soria Nieto}, {Souchay}, {Spagna}, {Spoto}, {Stampa}, {Steele},
  {Steidelm{\"u}ller}, {Stephenson}, {Stoev}, {Suess}, {S{\"u}veges}, {Surdej},
  {Szabados}, {Szegedi-Elek}, {Tapiador}, {Taris}, {Tauran}, {Taylor},
  {Teixeira}, {Terrett}, {Tingley}, {Trager}, {Turon}, {Ulla}, {Utrilla},
  {Valentini}, {van Elteren}, {Van Hemelryck}, {van Leeuwen}, {Varadi},
  {Vecchiato}, {Veljanoski}, {Via}, {Vicente}, {Vogt}, {Voss}, {Votruba},
  {Voutsinas}, {Walmsley}, {Weiler}, {Weingrill}, {Werner}, {Wevers},
  {Whitehead}, {Wyrzykowski}, {Yoldas}, {{\v{Z}}erjal}, {Zucker}, {Zurbach},
  {Zwitter}, {Alecu}, {Allen}, {Allende Prieto}, {Amorim},
  {Anglada-Escud{\'e}}, {Arsenijevic}, {Azaz}, {Balm}, {Beck}, {Bernstein},
  {Bigot}, {Bijaoui}, {Blasco}, {Bonfigli}, {Bono}, {Boudreault}, {Bressan},
  {Brown}, {Brunet}, {Bunclark}, {Buonanno}, {Butkevich}, {Carret}, {Carrion},
  {Chemin}, {Ch{\'e}reau}, {Corcione}, {Darmigny}, {de Boer}, {de Teodoro}, {de
  Zeeuw}, {Delle Luche}, {Domingues}, {Dubath}, {Fodor}, {Fr{\'e}zouls},
  {Fries}, {Fustes}, {Fyfe}, {Gallardo}, {Gallegos}, {Gardiol}, {Gebran},
  {Gomboc}, {G{\'o}mez}, {Grux}, {Gueguen}, {Heyrovsky}, {Hoar}, {Iannicola},
  {Isasi Parache}, {Janotto}, {Joliet}, {Jonckheere}, {Keil}, {Kim},
  {Klagyivik}, {Klar}, {Knude}, {Kochukhov}, {Kolka}, {Kos}, {Kutka}, {Lainey},
  {LeBouquin}, {Liu}, {Loreggia}, {Makarov}, {Marseille}, {Martayan},
  {Martinez-Rubi}, {Massart}, {Meynadier}, {Mignot}, {Munari}, {Nguyen},
  {Nordlander}, {Ocvirk}, {O'Flaherty}, {Olias Sanz}, {Ortiz}, {Osorio},
  {Oszkiewicz}, {Ouzounis}, {Palmer}, {Park}, {Pasquato}, {Peltzer}, {Peralta},
  {P{\'e}turaud}, {Pieniluoma}, {Pigozzi}, {Poels}, {Prat}, {Prod'homme},
  {Raison}, {Rebordao}, {Risquez}, {Rocca-Volmerange}, {Rosen}, {Ruiz-Fuertes},
  {Russo}, {Sembay}, {Serraller Vizcaino}, {Short}, {Siebert}, {Silva},
  {Sinachopoulos}, {Slezak}, {Soffel}, {Sosnowska}, {Strai{\v{z}}ys}, {ter
  Linden}, {Terrell}, {Theil}, {Tiede}, {Troisi}, {Tsalmantza}, {Tur},
  {Vaccari}, {Vachier}, {Valles}, {Van Hamme}, {Veltz}, {Virtanen}, {Wallut},
  {Wichmann}, {Wilkinson}, {Ziaeepour}, \& {Zschocke}}]{Gaia-2016}
{Gaia Collaboration}, {Prusti}, T., {de Bruijne}, J.~H.~J., {et~al.} 2016,
  \aap, 595, A1

\bibitem[{{Garc{\'{\i}}a} {et~al.}(2010){Garc{\'{\i}}a}, {Mathur}, {Salabert},
  {Ballot}, {R{\'e}gulo}, {Metcalfe}, \& {Baglin}}]{Garcia-2010}
{Garc{\'{\i}}a}, R.~A., {Mathur}, S., {Salabert}, D., {et~al.} 2010, Science,
  329, 1032

\bibitem[{{Gavras} {et~al.}(2023){Gavras}, {Rimoldini}, {Nienartowicz}, {de
  Fombelle}, {Holl}, {{\'A}brah{\'a}m}, {Audard}, {Carnerero}, {Clementini},
  {De Ridder}, {Distefano}, {Garcia-Lario}, {Garofalo}, {K{\'o}sp{\'a}l},
  {Kruszy{\'n}ska}, {Kun}, {Lecoeur-Ta{\"\i}bi}, {Marton}, {Mazeh}, {Mowlavi},
  {Raiteri}, {Ripepi}, {Szabados}, {Zucker}, \& {Eyer}}]{Gavras-2023}
{Gavras}, P., {Rimoldini}, L., {Nienartowicz}, K., {et~al.} 2023, \aap, 674,
  A22

\bibitem[{{Guenther} {et~al.}(2012){Guenther}, {Gandolfi}, {Sebastian},
  {Deleuil}, {Moutou}, \& {Cusano}}]{Guenther-2012}
{Guenther}, E.~W., {Gandolfi}, D., {Sebastian}, D., {et~al.} 2012, \aap, 543,
  A125

\bibitem[{{Hajdu} {et~al.}(2022){Hajdu}, {Mat{\'e}csa}, {Sallai}, \&
  {B{\'o}di}}]{Hajdu-2022}
{Hajdu}, T., {Mat{\'e}csa}, B., {Sallai}, J.~M., \& {B{\'o}di}, A. 2022,
  \mnras, 516, 5165

\bibitem[{{Hareter}(2012)}]{Hareter-2012}
{Hareter}, M. 2012, Astronomische Nachrichten, 333, 1048

\bibitem[{{Jetsu} \& {Pelt}(1996)}]{Jetsu-1996}
{Jetsu}, L. \& {Pelt}, J. 1996, \aaps, 118, 587

\bibitem[{{Lanza} {et~al.}(2009){Lanza}, {Pagano}, {Leto}, {Messina},
  {Aigrain}, {Alonso}, {Auvergne}, {Baglin}, {Barge}, {Bonomo}, {Boumier},
  {Collier Cameron}, {Comparato}, {Cutispoto}, {de Medeiros}, {Foing},
  {Kaiser}, {Moutou}, {Parihar}, {Silva-Valio}, \& {Weiss}}]{Lanza-2009}
{Lanza}, A.~F., {Pagano}, I., {Leto}, G., {et~al.} 2009, \aap, 493, 193

\bibitem[{{Lapeyrere} {et~al.}(2006){Lapeyrere}, {Bernardi}, {Buey},
  {Auvergne}, \& {Tiph{\`e}ne}}]{Lapeyrere-2006}
{Lapeyrere}, V., {Bernardi}, P., {Buey}, J.~T., {Auvergne}, M., \&
  {Tiph{\`e}ne}, D. 2006, \mnras, 365, 1171

\bibitem[{{Le{\~a}o} {et~al.}(2015){Le{\~a}o}, {Pasquini}, {Ferreira Lopes},
  {Neves}, {Valcarce}, {de Oliveira}, {Freire da Silva}, {de Freitas}, {Canto
  Martins}, {Janot-Pacheco}, {Baglin}, \& {De Medeiros}}]{Leao-2015}
{Le{\~a}o}, I.~C., {Pasquini}, L., {Ferreira Lopes}, C.~E., {et~al.} 2015,
  \aap, 582, A85

\bibitem[{{L{\'e}ger} {et~al.}(2009){L{\'e}ger}, {Rouan}, {Schneider}, {Barge},
  {Fridlund}, {Samuel}, {Ollivier}, {Guenther}, {Deleuil}, {Deeg}, {Auvergne},
  {Alonso}, {Aigrain}, {Alapini}, {Almenara}, {Baglin}, {Barbieri}, {Bruntt},
  {Bord{\'e}}, {Bouchy}, {Cabrera}, {Catala}, {Carone}, {Carpano}, {Csizmadia},
  {Dvorak}, {Erikson}, {Ferraz-Mello}, {Foing}, {Fressin}, {Gandolfi},
  {Gillon}, {Gondoin}, {Grasset}, {Guillot}, {Hatzes}, {H{\'e}brard}, {Jorda},
  {Lammer}, {Llebaria}, {Loeillet}, {Mayor}, {Mazeh}, {Moutou}, {P{\"a}tzold},
  {Pont}, {Queloz}, {Rauer}, {Renner}, {Samadi}, {Shporer}, {Sotin}, {Tingley},
  {Wuchterl}, {Adda}, {Agogu}, {Appourchaux}, {Ballans}, {Baron}, {Beaufort},
  {Bellenger}, {Berlin}, {Bernardi}, {Blouin}, {Baudin}, {Bodin}, {Boisnard},
  {Boit}, {Bonneau}, {Borzeix}, {Briet}, {Buey}, {Butler}, {Cailleau},
  {Cautain}, {Chabaud}, {Chaintreuil}, {Chiavassa}, {Costes}, {Cuna Parrho},
  {de Oliveira Fialho}, {Decaudin}, {Defise}, {Djalal}, {Epstein}, {Exil},
  {Faur{\'e}}, {Fenouillet}, {Gaboriaud}, {Gallic}, {Gamet}, {Gavalda},
  {Grolleau}, {Gruneisen}, {Gueguen}, {Guis}, {Guivarc'h}, {Guterman},
  {Hallouard}, {Hasiba}, {Heuripeau}, {Huntzinger}, {Hustaix}, {Imad},
  {Imbert}, {Johlander}, {Jouret}, {Journoud}, {Karioty}, {Kerjean},
  {Lafaille}, {Lafond}, {Lam-Trong}, {Landiech}, {Lapeyrere}, {Larqu{\'e}},
  {Laudet}, {Lautier}, {Lecann}, {Lefevre}, {Leruyet}, {Levacher}, {Magnan},
  {Mazy}, {Mertens}, {Mesnager}, {Meunier}, {Michel}, {Monjoin}, {Naudet},
  {Nguyen-Kim}, {Orcesi}, {Ottacher}, {Perez}, {Peter}, {Plasson}, {Plesseria},
  {Pontet}, {Pradines}, {Quentin}, {Reynaud}, {Rolland}, {Rollenhagen},
  {Romagnan}, {Russ}, {Schmidt}, {Schwartz}, {Sebbag}, {Sedes}, {Smit},
  {Steller}, {Sunter}, {Surace}, {Tello}, {Tiph{\`e}ne}, {Toulouse}, {Ulmer},
  {Vandermarcq}, {Vergnault}, {Vuillemin}, \& {Zanatta}}]{Leger-2009}
{L{\'e}ger}, A., {Rouan}, D., {Schneider}, J., {et~al.} 2009, A\&A, 506, 287

\bibitem[{{Lomb}(1976)}]{Lomb-1976}
{Lomb}, N.~R. 1976, \apss, 39, 447

\bibitem[{{Maciel} {et~al.}(2011){Maciel}, {Osorio}, \& {De
  Medeiros}}]{Maciel-2011}
{Maciel}, S.~C., {Osorio}, Y.~F.~M., \& {De Medeiros}, J.~R. 2011, \na, 16, 68

\bibitem[{{Marsakov} {et~al.}(2011){Marsakov}, {Koval'}, {Borkova}, \&
  {Shapovalov}}]{Marsakov-2011}
{Marsakov}, V.~A., {Koval'}, V.~V., {Borkova}, T.~V., \& {Shapovalov}, M.~V.
  2011, Astronomy Reports, 55, 667

\bibitem[{{Michel} {et~al.}(2017){Michel}, {Dupret}, {Reese}, {Ouazzani},
  {Debosscher}, {Hern{\'a}ndez}, {Belkacem}, {Samadi}, {Salmon}, {Suarez}, \&
  {Forteza}}]{Michel-2017}
{Michel}, E., {Dupret}, M.-A., {Reese}, D., {et~al.} 2017, in European Physical
  Journal Web of Conferences, Vol. 160, European Physical Journal Web of
  Conferences, 03001

\bibitem[{{Mislis} {et~al.}(2010){Mislis}, {Schmitt}, {Carone}, {Guenther}, \&
  {P{\"a}tzold}}]{Mislis-2010}
{Mislis}, D., {Schmitt}, J.~H.~M.~M., {Carone}, L., {Guenther}, E.~W., \&
  {P{\"a}tzold}, M. 2010, \aap, 522, A86

\bibitem[{{Moya} {et~al.}(2017){Moya}, {Su{\'a}rez}, {Garc{\'\i}a
  Hern{\'a}ndez}, \& {Mendoza}}]{Moya-2017}
{Moya}, A., {Su{\'a}rez}, J.~C., {Garc{\'\i}a Hern{\'a}ndez}, A., \& {Mendoza},
  M.~A. 2017, \mnras, 471, 2491

\bibitem[{{Nikzat} {et~al.}(2022){Nikzat}, {Ferreira Lopes}, {Catelan},
  {Contreras Ramos}, {Zoccali}, {Rojas-Arriagada}, {Braga}, {Minniti},
  {Borissova}, \& {Becker}}]{Nikzat-2022}
{Nikzat}, F., {Ferreira Lopes}, C.~E., {Catelan}, M., {et~al.} 2022, \aap, 660,
  A35

\bibitem[{{Ollivier} {et~al.}(2016){Ollivier}, {Tiph{\`e}ne}, {Samadi},
  {Levacher}, \& {CoRot Team}}]{Ollivier-2016}
{Ollivier}, M., {Tiph{\`e}ne}, D., {Samadi}, R., {Levacher}, P., \& {CoRot
  Team}. 2016, {V.2 CoRoT heritage in future missions}, 237

\bibitem[{{Paltani}(2004)}]{Paltani-2004}
{Paltani}, S. 2004, \aap, 420, 789

\bibitem[{{Papageorgiou} {et~al.}(2018){Papageorgiou}, {Catelan},
  {Christopoulou}, {Drake}, \& {Djorgovski}}]{Papageorgiou-2018}
{Papageorgiou}, A., {Catelan}, M., {Christopoulou}, P.-E., {Drake}, A.~J., \&
  {Djorgovski}, S.~G. 2018, \apjs, 238, 4

\bibitem[{Pedregosa {et~al.}(2011)Pedregosa, Varoquaux, Gramfort, Michel,
  Thirion, Grisel, Blondel, Prettenhofer, Weiss, Dubourg,
  {et~al.}}]{Pedregosa2011scikit-learn}
Pedregosa, F., Varoquaux, G., Gramfort, A., {et~al.} 2011, Journal of Machine
  Learning Research, 12, 2825

\bibitem[{{Pr{\v s}a} {et~al.}(2011){Pr{\v s}a}, {Batalha}, {Slawson}, {Doyle},
  {Welsh}, {Orosz}, {Seager}, {Rucker}, {Mjaseth}, {Engle}, {Conroy},
  {Jenkins}, {Caldwell}, {Koch}, \& {Borucki}}]{Prsa-2011}
{Pr{\v s}a}, A., {Batalha}, N., {Slawson}, R.~W., {et~al.} 2011, \aj, 141, 83

\bibitem[{{Queloz} {et~al.}(2009){Queloz}, {Bouchy}, {Moutou}, {Hatzes},
  {H{\'e}brard}, {Alonso}, {Auvergne}, {Baglin}, {Barbieri}, {Barge}, {Benz},
  {Bord{\'e}}, {Deeg}, {Deleuil}, {Dvorak}, {Erikson}, {Ferraz Mello},
  {Fridlund}, {Gandolfi}, {Gillon}, {Guenther}, {Guillot}, {Jorda}, {Hartmann},
  {Lammer}, {L{\'e}ger}, {Llebaria}, {Lovis}, {Magain}, {Mayor}, {Mazeh},
  {Ollivier}, {P{\"a}tzold}, {Pepe}, {Rauer}, {Rouan}, {Schneider},
  {Segransan}, {Udry}, \& {Wuchterl}}]{Queloz-2009}
{Queloz}, D., {Bouchy}, F., {Moutou}, C., {et~al.} 2009, \aap, 506, 303

\bibitem[{{Ratcliffe} {et~al.}(2023){Ratcliffe}, {Minchev}, {Anders},
  {Khoperskov}, {Guiglion}, {Buck}, {Cunha}, {Queiroz}, {Nitschelm},
  {Meszaros}, {Steinmetz}, {de Jong}, {Nepal}, {Lane}, \&
  {Sobeck}}]{Ratcliffe-2023}
{Ratcliffe}, B., {Minchev}, I., {Anders}, F., {et~al.} 2023, \mnras
  [\eprint[arXiv]{2305.13378}]

\bibitem[{{Ricker} {et~al.}(2015){Ricker}, {Winn}, {Vanderspek}, {Latham},
  {Bakos}, {Bean}, {Berta-Thompson}, {Brown}, {Buchhave}, {Butler}, {Butler},
  {Chaplin}, {Charbonneau}, {Christensen-Dalsgaard}, {Clampin}, {Deming},
  {Doty}, {De Lee}, {Dressing}, {Dunham}, {Endl}, {Fressin}, {Ge}, {Henning},
  {Holman}, {Howard}, {Ida}, {Jenkins}, {Jernigan}, {Johnson}, {Kaltenegger},
  {Kawai}, {Kjeldsen}, {Laughlin}, {Levine}, {Lin}, {Lissauer}, {MacQueen},
  {Marcy}, {McCullough}, {Morton}, {Narita}, {Paegert}, {Palle}, {Pepe},
  {Pepper}, {Quirrenbach}, {Rinehart}, {Sasselov}, {Sato}, {Seager},
  {Sozzetti}, {Stassun}, {Sullivan}, {Szentgyorgyi}, {Torres}, {Udry}, \&
  {Villasenor}}]{Ricker-2015}
{Ricker}, G.~R., {Winn}, J.~N., {Vanderspek}, R., {et~al.} 2015, Journal of
  Astronomical Telescopes, Instruments, and Systems, 1, 014003

\bibitem[{{Sarro} {et~al.}(2013){Sarro}, {Debosscher}, {Neiner},
  {Bello-Garc{\'{\i}}a}, {Gonz{\'a}lez-Marcos}, {Prendes-Gero}, {Ordieres},
  {Le{\'o}n}, {Aerts}, \& {de Batz}}]{Sarro-2013}
{Sarro}, L.~M., {Debosscher}, J., {Neiner}, C., {et~al.} 2013, \aap, 550, A120

\bibitem[{{Scargle}(1982)}]{Scargle-1982}
{Scargle}, J.~D. 1982, \apj, 263, 835

\bibitem[{{Sebastian} {et~al.}(2022){Sebastian}, {Guenther}, {Deleuil},
  {Dorsch}, {Heber}, {Heuser}, {Gandolfi}, {Grziwa}, {Deeg}, {Alonso},
  {Bouchy}, {Csizmadia}, {Cusano}, {Fridlund}, {Geier}, {Irrgang}, {Korth},
  {Nespral}, {Rauer}, {Tal-Or}, \& {CoRoT-team}}]{Sebastian-2022}
{Sebastian}, D., {Guenther}, E.~W., {Deleuil}, M., {et~al.} 2022, \mnras, 516,
  636

\bibitem[{van~der Maaten \& Hinton(2008)}]{vanDerMaaten2008}
van~der Maaten, L. \& Hinton, G. 2008, Journal of Machine Learning Research, 9,
  2579

\bibitem[{{Walkowicz} \& {Basri}(2013)}]{Walkowicz-2013}
{Walkowicz}, L.~M. \& {Basri}, G.~S. 2013, \mnras, 436, 1883

\bibitem[{{Watson} {et~al.}(2014){Watson}, {Henden}, \& {Price}}]{Watson-2014}
{Watson}, C., {Henden}, A.~A., \& {Price}, A. 2014, VizieR Online Data Catalog,
  1, 2027

\bibitem[{{Wenger} {et~al.}(2000){Wenger}, {Ochsenbein}, {Egret}, {Dubois},
  {Bonnarel}, {Borde}, {Genova}, {Jasniewicz}, {Lalo{\"e}}, {Lesteven}, \&
  {Monier}}]{simbad-2000}
{Wenger}, M., {Ochsenbein}, F., {Egret}, D., {et~al.} 2000, \aaps, 143, 9

\bibitem[{{Zechmeister} \& {K{\"u}rster}(2009)}]{Zechmeister-2009}
{Zechmeister}, M. \& {K{\"u}rster}, M. 2009, \aap, 496, 577

\bibitem[{Zhang \& Wang(2006)}]{Zhang2006MLLEML}
Zhang, Z. \& Wang, J. 2006, in NIPS

\bibitem[{Zhou {et~al.}(2003)Zhou, Bousquet, Lal, Weston, \&
  Sch{\"o}lkopf}]{Zhou2003LearningWL}
Zhou, D., Bousquet, O., Lal, T.~N., Weston, J., \& Sch{\"o}lkopf, B. 2003, in
  NIPS

\bibitem[{{Zoccali} {et~al.}(2017){Zoccali}, {Vasquez}, {Gonzalez}, {Valenti},
  {Rojas-Arriagada}, {Minniti}, {Rejkuba}, {Minniti}, {McWilliam}, {Babusiaux},
  {Hill}, \& {Renzini}}]{Zoccali-2017}
{Zoccali}, M., {Vasquez}, S., {Gonzalez}, O.~A., {et~al.} 2017, \aap, 599, A12

\bibitem[{{Zorec} {et~al.}(2023){Zorec}, {Hubert}, {Martayan}, \&
  {Fr{\'e}mat}}]{Zorec-2023}
{Zorec}, J., {Hubert}, A.~M., {Martayan}, C., \& {Fr{\'e}mat}, Y. 2023, \aap,
  676, A81

\end{thebibliography}
\end{document}